\documentclass[twocolumn]{aastex}
\usepackage{url}
\usepackage{graphicx}

\begin{document}

\newcommand{\Ktwo}{{\em K2 }}
\newcommand{\Ktwoo}{{\em K2}}
\newcommand{\Kepler}{{\em Kepler }}
\newcommand{\Keplerr}{{\em Kepler}}
\newcommand{\ptfeb}{PTFEB132.707+19.810 }
\newcommand{\ptfebb}{PTFEB132.707+19.810}

\title{The Factory and the Beehive III: \\ \ptfebb, A Low-Mass Eclipsing Binary in Praesepe Observed by PTF and \Ktwo}

\author{
Adam L. Kraus\altaffilmark{1},
Stephanie T. Douglas\altaffilmark{2},
Andrew W. Mann\altaffilmark{1,3},
Marcel A. Ag\"ueros\altaffilmark{2},
Nicholas M. Law\altaffilmark{4},\\
Kevin R. Covey\altaffilmark{5},
Gregory A. Feiden\altaffilmark{6}
Aaron C. Rizzuto\altaffilmark{1},
Andrew W. Howard\altaffilmark{7},
Howard Isaacson\altaffilmark{8},\\
Eric Gaidos\altaffilmark{9},
Guillermo Torres\altaffilmark{10},
Gaspar Bakos\altaffilmark{11,12}
}

\altaffiltext{1}{Department of Astronomy, The University of Texas at Austin, Austin, TX 78712, USA}
\altaffiltext{2}{Columbia University, Department of Astronomy, 550 West 120th Street, New York, NY 10027, USA}
\altaffiltext{3}{Hubble Fellow}
\altaffiltext{4}{Department of Physics and Astronomy, University of North Carolina, Chapel Hill, NC 27599, USA}
\altaffiltext{5}{Western Washington University, Department of Physics \& Astronomy, Bellingham, WA 98225, USA}
\altaffiltext{6}{Department of Physics, University of North Georgia, Dahlonega, GA 30597, USA}
\altaffiltext{7}{California Institute of Technology, 1200 E California Boulevard, Pasadena, CA 91125, USA}
\altaffiltext{8}{University of California, Berkeley, Berkeley, CA 94720, USA}
\altaffiltext{9}{Department of Geology and Geophysics, University of Hawaii at Manoa, Honolulu, HI 96822, USA}
\altaffiltext{10}{Harvard-Smithsonian Center for Astrophysics, 60 Garden St., Cambridge, MA 02138, USA}
\altaffiltext{11}{Department of Astrophysical Sciences, Princeton University, Princeton, NJ 08544, USA}
\altaffiltext{12}{Packard Fellow}

\begin{abstract}

Theoretical models of stars constitute a fundamental bedrock upon which much of astrophysics is built, but large swaths of model parameter space remain uncalibrated by observations. The best calibrators are eclipsing binaries in clusters, allowing measurement of masses, radii, luminosities, and temperatures, for stars of known metallicity and age. We present the discovery and detailed characterization of \ptfebb, a $P=6.0$ day eclipsing binary in the Praesepe cluster ($\tau\sim600$--800~Myr; [Fe/H]$=0.14\pm0.04$). The system contains two late-type stars (SpT$_P$=M3.5$\pm$0.2; SpT$_S$=M4.3$\pm$0.7) with precise masses ($M_p=0.3953\pm0.0020$~$M_{\odot}$; $M_s=0.2098\pm0.0014$~$M_{\odot}$) and radii ($R_p=0.363\pm0.008$~$R_{\odot}$; $R_s=0.272\pm0.012$~$R_{\odot}$). Neither star meets the predictions of stellar evolutionary models. The primary has the expected radius, but is cooler and less luminous, while the secondary has the expected luminosity, but is cooler and substantially larger (by 20\%). The system is not tidally locked or circularized. Exploiting a fortuitous 4:5 commensurability between $P_{orb}$ and $P_{rot,prim}$, we demonstrate that fitting errors from the unknown spot configuration only change the inferred radii by $\la$1--2\%. We also analyze subsets of data to test the robustness of radius measurements; the radius sum is more robust to systematic errors and preferable for model comparisons. We also test plausible changes in limb darkening, and find corresponding uncertainties of $\sim$1\%. Finally, we validate our pipeline using extant data for GU Boo, finding that our independent results match previous radii to within the mutual uncertainties (2--3\%). We therefore suggest that the substantial discrepancies are astrophysical; since they are larger than for old field stars, they may be tied to the intermediate age of \ptfebb.

\end{abstract}

\keywords{stars:binaries:eclipsing, stars:individual(PTFEB132.707+19.810), stars:fundamental parameters, stars:evolution, stars:low-mass, stars:starspots}

\section{Introduction} \label{sec:intro}

The fundamental properties of stars establish the foundation for much of astrophysics, and uncertainties in model-derived stellar properties propagate into systematic uncertainties in the initial mass function \citep{Bastian:2010wb}, the age-activity-rotation relation \citep{Agueros:2011dp,Douglas:2016fk,Rebull:2016qy}, and exoplanet masses and radii (e.g., \citealt{Gaidos:2012fk,Bastien:2014lr}). The relations between mass, radius, luminosity, temperature, and metallicity, or subsets thereof, have traditionally been calibrated with observations of the Sun, stellar populations, or binary systems. However, these calibrations remain sparse and observationally expensive for low-mass stars due to their intrinsically low luminosities. The past 15 years have seen benchmark calibrations for the relations for mass-luminosity (with visual binaries; \citealt{Delfosse:2000it}), luminosity-temperature-radius (with long-baseline interferometry; \citealt{Boyajian:2012il}), and luminosity-temperature-metallicity (with spectroscopy; \citealt{Mann:2015ys}), typically with a scatter of $\sim$5\% for mass or radius measurements.

Direct measurements of the mass-radius relation are typically derived from studies of field eclipsing binary systems (e.g., \citealt{Torres:2010lr}), for which the orbit yields masses and the eclipses yield radii. Eclipsing binaries can yield mass and radius measurements with formal uncertainties of $\la1\%$, surpassing the precision offered by visual binaries or long-baseline interferometry, but due to their faintness only a modest number of systems with low-mass components ($M_p \la 0.7 M_{\odot}$) have been discovered and characterized. The systems discovered to date have suggested that models and observations are discrepant in at least some cases, with models predicting radii that are up to 10\% too low for a given mass (e.g., \citealt{Lopez-Morales:2007kz,Torres:2010lr}). This problem has been seen for solar-type stars for several decades (e.g., \citealt{Popper:1997aa,Torres:2006dk,Torres:2008aa,Clausen:2009aa}. The most recent compilations by \citet{Dittmann:2017aa} and \citet{Iglesias-Marzoa:2017aa} show that this scatter exists for the entire mass range from $0.2 < M < 1.0 M_{\odot}$; some stellar radii match model predictions, but most stars are larger. 

It remains unclear whether the radius discrepancy results from physics common to all low-mass stellar interiors (e.g., requiring extra opacities or modified treatments of convection, metallicity, or magnetic fields; \citealt{Feiden:2014yu,Feiden:2014qf}), or a systematic effect specific to the short-period binary systems that are most likely to eclipse. These short-period systems  often tidally locked to rapid rotation periods that sustain high levels of magnetic activity, possibly leading to larger radii than are predicted by non-magnetic stellar models \citep{Chabrier:2007cs,Morales:2009rb,Kraus:2011lr}. However, even some very long-period systems are larger than models would predict \citep{Irwin:2011ys}.

Eclipsing binaries in star clusters offer even stronger tests, allowing measurement of the masses, radii, luminosities, temperatures, metallicities, and ages of the component stars, but those systems have been rare to date. \ptfeb - also AD 3814 \citep{Adams:2002sy}, 2MASS J08504984+1948364 \citep{Cutri:2003jh}, and EPIC 211972086 \citep{Huber:2016zr} - was first suggested as a candidate member of the Praesepe open cluster by \citet{Adams:2002sy}, who assessed a membership probability of 48.4\% based on its 2MASS/POSS colors and proper motion. The membership probability was subseqently revised upward to 97.9\% by \citet{Kraus:2007mz} based on its 2MASS/USNO-B1.0/SDSS photometry and proper motion, while \citet{Boudreault:2012fk} assessed a membership probability of 86\% based on UKIDSS photometry and proper motion. \citet{Kraus:2007mz} estimated the spectral type to be M3.4 based on the broadband optical/NIR SED, and \citet{West:2011zl} analyzed the SDSS spectrum to estimate a spectral type of M5. 

\ptfeb was closely inspected by the PTF Open Cluster Survey (POCS) due to its likely Praesepe membership and its nature as a mid-M dwarf. \citet{Agueros:2011dp} measured a rotation period of $P_{rot} = 7.43$ days based on clear periodic brightness variations. As we describe below, \ptfeb also was identified to be an eclipsing binary with an orbital period of $P_{orb} = 6.0$ days that has not locked its stars into synchronous rotation. The star has otherwise remained anonymous in the literature until this year, when its eclipsing nature was recognized by others studying \Ktwo data in Praesepe \citep{Rebull:2017aa,Douglas:2017aa,Gillen:2017aa}. In this paper, we analyze our discovery light curve from PTF, followup spectroscopic observations, and the newly released \Ktwo light curve of \ptfeb in order to measure the stellar properties of this mid-M-type eclipsing binary system and test the predictions of stellar evolutionary models. In Section~\ref{sec:obs}, we describe our observations of this system, and in Section~\ref{sec:analysis} we describe the analysis used to interpret those observations. We summarize the resulting properties for the system in Section~\ref{sec:results}, as well as testing the robustness of our results. Finally, in Section~\ref{sec:models} we discuss the implications of our results for the current generation of stellar evolutionary models.

\section{Observations} \label{sec:obs}

\subsection{Palomar Transient Factory Photometry} \label{sec:obs:ptf}

The Palomar Transient Factory (PTF) uses wide-field photometric observations from the robotic 48 inch Samuel Oschin telescope (hereafter P48), a Schmidt camera with a wide focal plane. When \ptfeb was observed in 2010 and 2011, P48 was equipped with the CFH12K mosaic camera that had been installed on the Canada-France-Hawaii Telescope \citep{Rahmer:2008uq}. The camera covered a total field of 8 deg$^2$ (with 7.26 deg$^2$ of useful area since one chip is nonfunctional), sampled with a pixel scale of 1\arcsec. The observations that we report were taken in the standard PTF observing mode, which uses 60s integrations through a Mould $R$ filter, yielding photometry for all stars down to $m_R \sim 20$ mag \citep{Ofek:2012eu}. The Praesepe field was typically observed 1--2 times per night as part of the POCS campaign \citep{Agueros:2011dp}, but it was also observed at a more rapid cadence (every 15 minutes, yielding 15--30 images per night) on some nights as part of the PTF/M-dwarfs campaign \citep{Law:2011jk,Law:2012qa}. \ptfeb was observed 622 times over the course of these two PTF programs.

The construction and analysis of light curves for the POCS and PTF-M-dwarfs programs were described in more detail by \citet{Law:2009rt,Law:2012qa}, and \citet{Agueros:2011dp}. To briefly summarize, the images were first calibrated to correct for cross-talk, perform bias/overscan subtraction, and divide by a contemporaneous superflat. The data were then processed with SExtractor to measure source photometry, and sources were matched across all epochs. The photometric zero points for each epoch were initially estimated based on SDSS photometry for bright stars in the field, and then the zero point for each epoch was optimized to minimize the median photometric variability of all remaining sources, rejecting apparently variable sources. The pipeline typically achieved a photometric stability of 3--5 mmag over periods of months; the photometric uncertainties for all observations of \ptfeb were limited by photon noise.

\ptfeb was identified as a candidate eclipsing binary system using a Box Least Squares algorithm \citep{Kovacs:2002fj} that phased the light curves to all possible periods and searched for a transit- or eclipse-like signature. The identification of eclipses was then visually confirmed.

\begin{figure*}
 \epsscale{0.90}
 \plotone{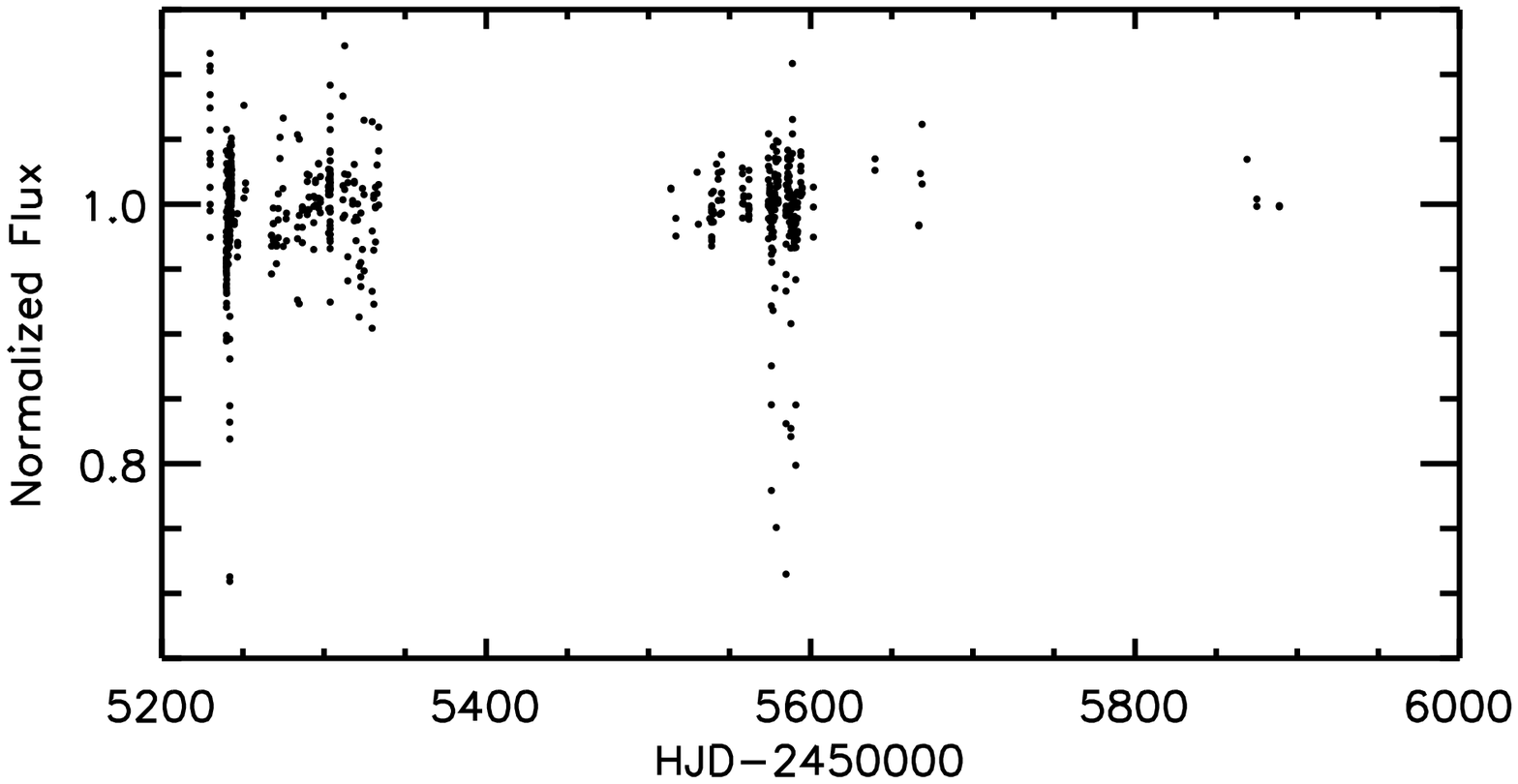}
 \caption{PTF Aperture photometry results for the Praesepe eclipsing binary \ptfebb, with fluxes normalized to the median value ($m_R = 17.03$ mag). \label{fig:ptfcurve}}
  \end{figure*}

We list all of the photometric measurements for \ptfeb in Table~\ref{tab:ptfphot}, and in Figure~\ref{fig:ptfcurve} we show the light curve spanning three observing seasons.

\subsection{\Ktwo Photometry} \label{sec:obs:k2}

\begin{figure*}
 \epsscale{0.8}
\includegraphics[scale=0.4,trim=0.0cm -3.5cm 0.0cm 0.0cm]{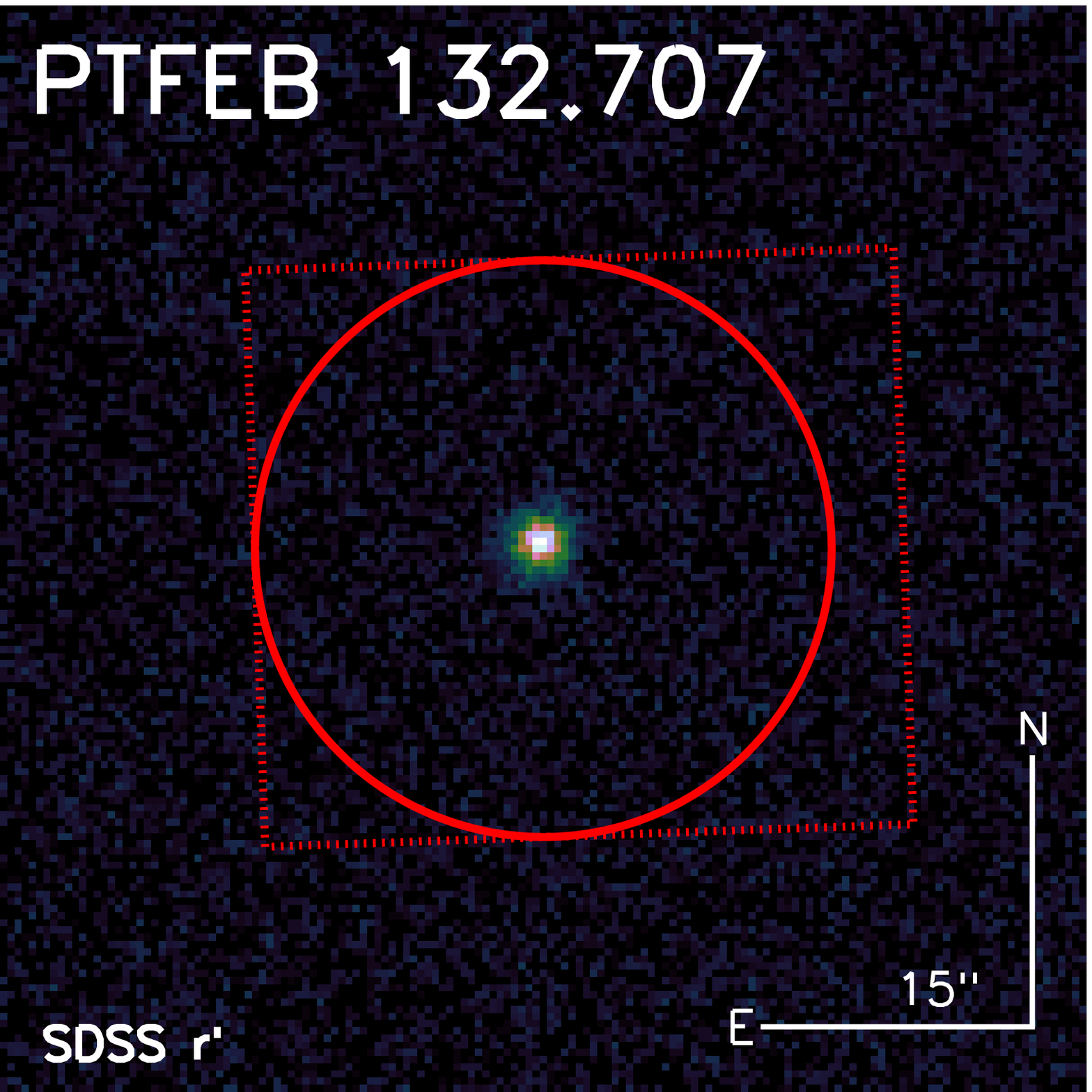} 
\includegraphics[scale=0.75,trim=0.0cm 0.0cm 0.0cm 0.0cm]{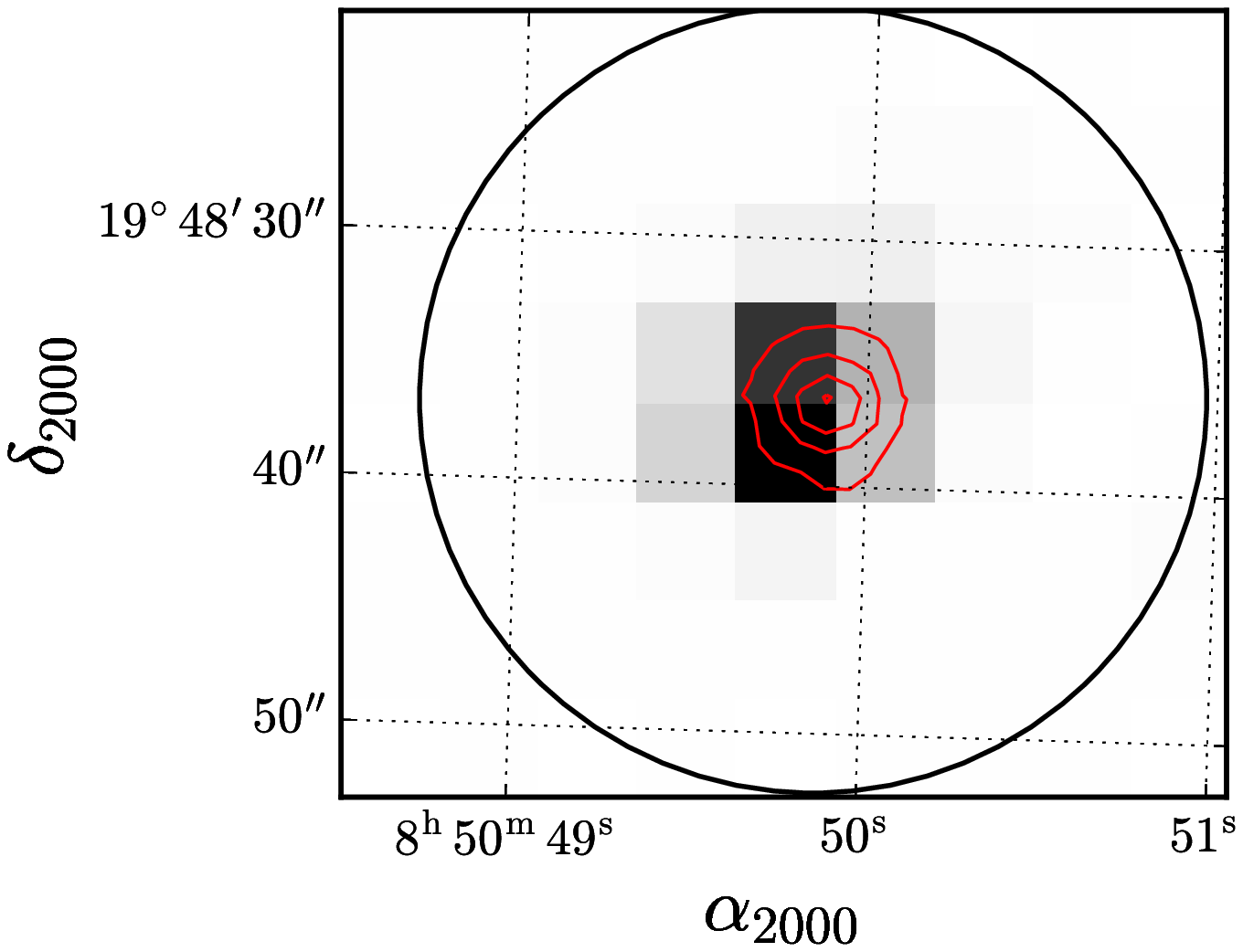}
 \caption{Left: SDSS $r$ postage stamp of \ptfeb (FOV=60\arcsec, North=Up) showing the \Ktwo postage stamp (red dotted box) and our adopted 4-pixel photometric aperture (red circle). The image is shown in a square-root stretch using the CubeHelix color palette \citep{Green:2011lr}. There are no nearby sources on the postage stamp, or anywhere in this image. \ptfeb has not been observed with adaptive optics, so there are no limits on closer companions, though the absence of a third set of spectral lines in our spectra suggests that there are no additional objects within $\Delta R < $2. Right:  Postage stamp of \ptfeb that was downloaded as part of \Ktwoo's Campaign 5, showing the coadded sum of all frames. The black circle shows the 4 pixel photometric aperture used in our analysis. The red contours show the flux distribution of the SDSS image.\label{fig:stamp}}
  \end{figure*}

\ptfeb was observed as EPIC 211972086 by the \Kepler spacecraft during Campaign 5 of its repurposed \Ktwo mission \citep{Howell:2014db}, for which it was proposed as a target by eight proposals, including ours (GO5095, PI Ag\"ueros). \Ktwo observed \ptfeb in long-cadence mode ($t_{int} = 29.4$ min) for 73.9 continuous days spanning 2015 April 27 to 2015 July 10, yielding 3402 exposures of the $9 \times 8$ pixel postage stamp centered on the target. We show the postage stamp superimposed on SDSS images of the field in Figure~\ref{fig:stamp} (left) and one frame from the K2 postage stamp in Figure~\ref{fig:stamp} (right). The \Ktwo data were downloaded from the Mikulski Archive for Space Telescopes as target pixel files, which contain the barycentric corrected observation times, the flux measured in each pixel at each epoch, and quality flags. We omitted the 15 exposures where the quality flag is not 0.

The \Kepler spacecraft pointing drifted by $\sim$1 pixel per 6 hours due to solar pressure, and the spacecraft executed a thruster fire every 6 hours to return to its optimal orientation. The data therefore contain a 6 hour periodicity with amplitude $\pm$0.5\% due to the star being sampled across the detector pixels (and their response function) differently over time \citep{Vanderburg:2014rr}. The light curve also show visually obvious sinusoidal periodicity with period $P_{rot} \sim 7.5$ days and amplitude $\pm 3\%$, as well as clear primary and secondary eclipses with a periodicity of $P_{orb} \sim 6.0$ days. The primary star flux contributed $\sim$75\% of the total optical flux (Section~\ref{sec:obs:hires}), so the origin of the sinusoidal periodicity was most likely rotational modulation due to spots on the primary. We prepared the light curve for eclipse fitting by using the methods described by \citet{Douglas:2016fk} to measure photometry and rectify the stellar and instrumental variability.

We began by computing and subtracting the background in each exposure using an iterative 3-sigma-clipped median, and then measured the flux of \ptfeb through a soft-edged circular aperture with a radius of 4 pixels, yielding a raw light curve. The aperture was centered on the photocenter in each exposure, so it tracked the drift of the target. We then detrended the long-timescale variability using the Python routine {\em supersmoother}\footnote{\url{https://github.com/jakevdp/supersmoother}} \citep{Friedman:1984aa} with a high bass-enhancement value ($\alpha = 10$).

We measured the period of the sinusoidal variability using the Lomb-Scargle function in the Python package {\em gatspy}, an implementation of the FFT-based algorithm from \citet{Press:1989fr}. We computed the periodogram power for $3 \times 10^4$ periods spanning 0.1--70.8 days and established false alarm probabilities using non-parametric bootstrap resampling to generate $10^3$ simulated light curves. To divide out the sinusoidal variability, we folded the light curve on the most likely period and used {\em supersmoother} to produce a smoothed periodic light curve. After iterating this procedure six times to remove stellar variability, we next measured the time-dependent flatfield at each epoch by calculating the median flux for the 21 other epochs with the closest centroid positions (in detector coordinates). We then returned to the raw light curve and first applied the flatfield correction, then fit and rectified the stellar variability. Finally, we rectified the remaining low-order power in the light curve by dividing the flux at each epoch by the median of all other non-eclipse fluxes observed within $\pm$12 hours.

\begin{figure*}
 \epsscale{0.90}
 \plotone{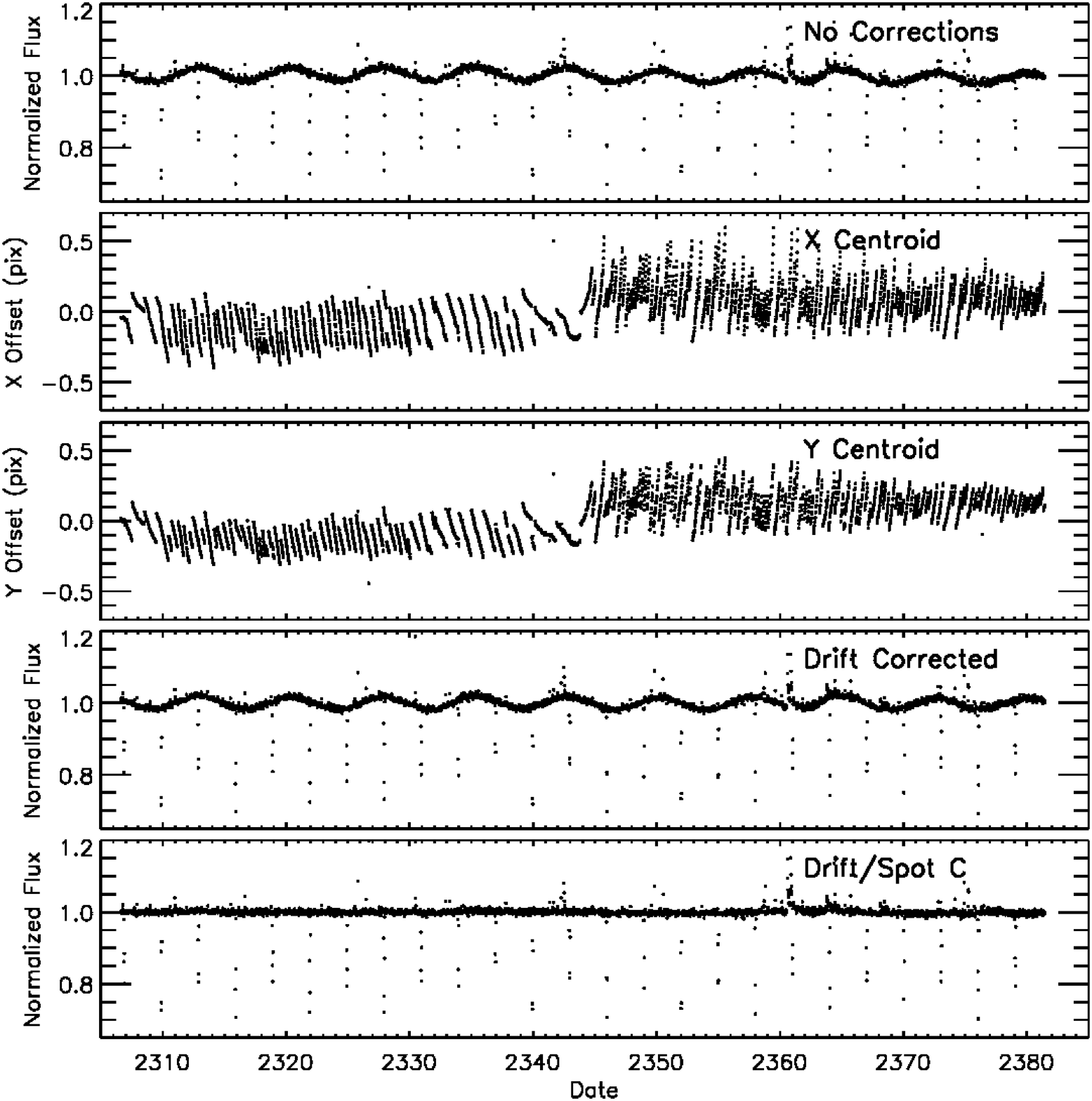}
 \caption{\Ktwo Aperture photometry results for the Praesepe eclipsing binary \ptfeb (EPIC 211972086). Time is specified in units of BJD-2454833, the standard time system for \Ktwoo. Top panel: Normalized light curve extracted from aperture photometry, without any subsequent detrending. Second and third panels: X and Y centroid positions, in pixels, as a function of time. The six-hour interval between thruster firings (which reset the telescope position) is evident in the positions, and the position information can be used to detrend flux variations as the target moves across the detector. Fourth panel: Normalized light curve after correcting flatfield variations due to telescope drift. Bottom panel: Normalized light curve after detrending the phase-folded mean stellar variability due to spots; this light curve is used for eclipse fitting. \label{fig:k2curve}}
  \end{figure*}

We show the stages of this process in Figure~\ref{fig:k2curve}, and present the resulting normalized fluxes in Table~\ref{tab:k2phot}. We found a final best-fit rotational period (most likely for the primary) of $P_{rot} = 7.46$ days, which is nearly identical to the period of $P = 7.43$ days measured by \citet{Agueros:2011dp}. We also repeated our analysis for the same light curve extracted by K2SC \citep{Aigrain:2016xy} and found a period of $P = 7.49$ days. The uncertainties in rotational periods are dominated by systematic effects due to evolution of the (unmeasurable) spot configuration, but different surveys of rotation in open clusters have typically yielded values that agree to within $\la$2--3\% (e.g., \citealt{Douglas:2016fk}), so these measurements are statistically indistinguishable. The light curve also contains numerous flares; one primary eclipse (at epoch 2361) and one secondary eclipse (at epoch 2364) were contaminated by flares, so we have omitted those observations from our light curve fits.

\subsection{Archival Photometry} \label{sec:obs:arcphot}

As part of our analysis to compute a bolometric flux for this system, we also have compiled all of the available (component unresolved) photometry in all-sky surveys. As we summarize in Table~\ref{tab:catphot}, we have used photometry from SDSS-DR9 ($u$, $g$, $r$, $i$, $z$; \citealt{Ahn:2012ys}), 2MASS ($J$,$H$,$K_s$; \citealt{Cutri:2003jh}), and AllWISE ($W1$, $W2$, $W3$; \citealt{Cutri:2013lq}).

\subsection{Keck/HIRES High-Resolution Spectroscopy and Radial Velocities} \label{sec:obs:hires}

After our discovery of eclipses in this system in 2010, we began a spectroscopic monitoring campaign to measure radial velocities (RVs) for the components. We obtained 20 high-dispersion spectra for the system using Keck-I and the HIRES spectrograph \citep{Vogt:1994ty}, which is a single-slit echelle spectrograph permanently mounted on the Nasmyth platform. Ten spectra were obtained using classical observing on three nights. These observations were performed using the red channel and C1 decker, and spanned a wavelength range of 4500--8900 \AA, yielding a spectral resolution of $R \sim 48$,000. We processed our HIRES data using the standard pipeline MAKEE\footnote{\url{http://www.astro.caltech.edu/~tb/makee/}}, which automatically extracts, flat-fields, and wavelength-calibrates spectra taken in most standard HIRES configurations. Ten additional spectra were obtained in a queue mode via collaboration with the California Planet Search (CPS) on nine nights, so as to obtain broader phase coverage for the system. These observations were performed using the red channel and C2 decker, and spanned a wavelength range of 3400--8100 \AA, also yielding $R \sim 48$,000. These data were extracted using the standard CPS pipeline \citep{Howard:2010kx}. In both cases, we refined the wavelength solution by cross-correlating the 7600 \AA\, O$_2$ telluric absorption band against that of the spectrophotometric standard star HZ 44 \citep{Massey:1988pt}.

Our analysis of the HIRES data to measure RVs is identical to the methods described in \cite{Kraus:2011lr,Kraus:2014rt,Kraus:2015mz}. For each spectrum, we measured the broadening function \citep{Rucinski:1999yr}\footnote{\url{http://www.astro.utoronto.ca/~rucinski/SVDcookbook.html}} with respect to three Keck/HIRES observations of two standard stars with similar temperature and metallicity: one observation of Gl 447, and two separate observations of Gl 83.1 on different nights. The broadening function is a better representation of the rotational broadening convolution than a cross-correlation, since it is less subject to ``peak pulling'' and produces a flatter continuum. We fit each broadening function with two Gaussian functions to determine the absolute primary and secondary star RVs ($v_p$ and $v_s$), the standard deviations of the line width due to rotation and instrumental resolution ($\sigma_{v_p}$ and $\sigma_{v_s}$), and the average flux ratio across all well-fit orders (which is estimated from the ratio of areas for the two peaks of the broadening function). We list these measurements in Table~\ref{tab:hiresRV}.

\begin{figure*}
 \includegraphics[scale=0.90,trim=0cm 0cm 0cm 0cm]{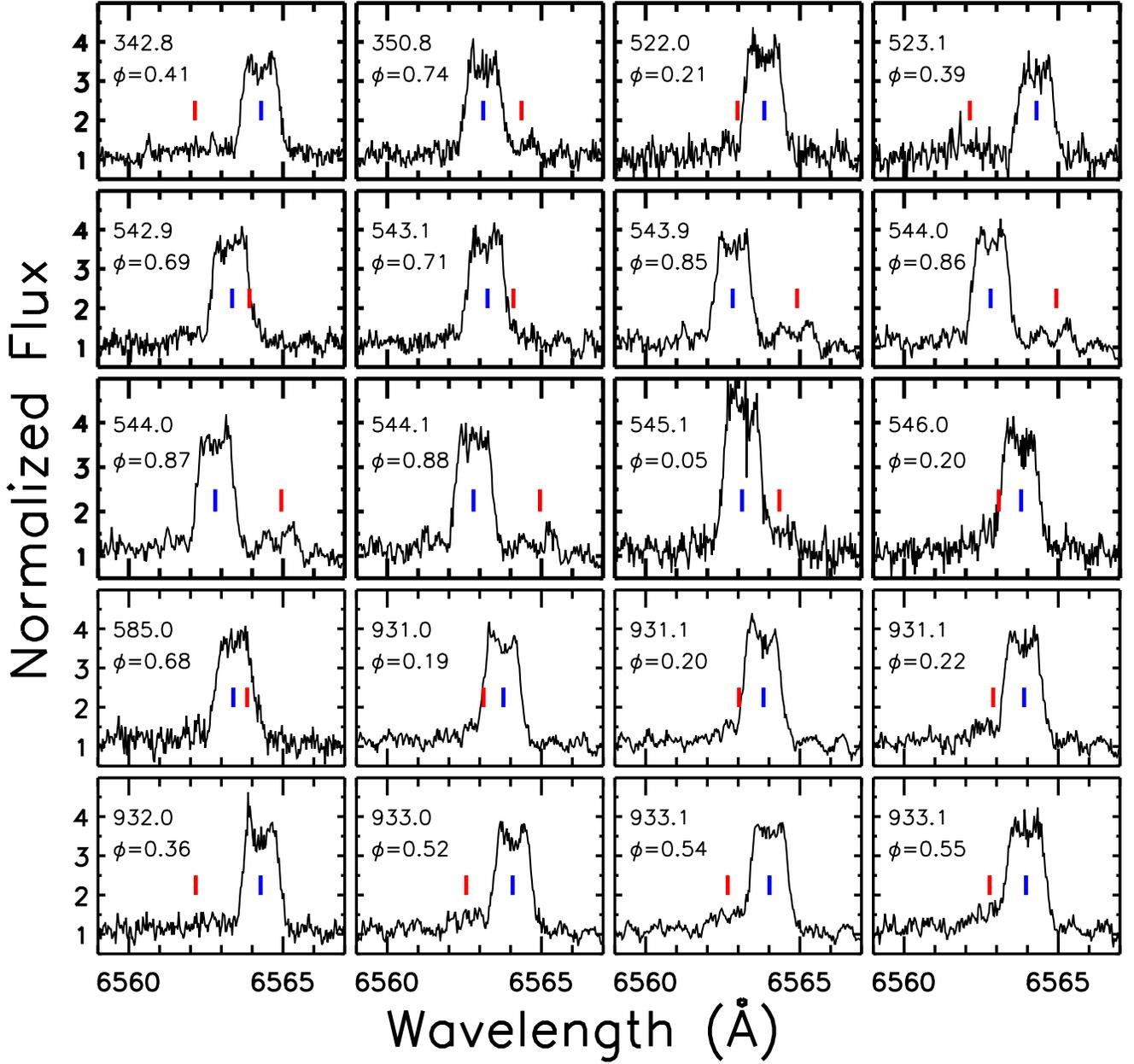}
 \caption{The spectral window around H$\alpha$ for the 20 Keck/HIRES spectra of \ptfebb. All spectra are normalized to unity in the continuum, and the wavelength scales are shifted to account for the heliocentric correction. The blue and red vertical lines show the expected wavelength of H$\alpha$ for the primary and secondary star respectively. The numbers list the epoch (HJD - 2455000) and the orbital phase. Emission is clearly visible for the primary star at all epochs, but it is only marginally detected for the secondary star. The H$\alpha$ emission lines are broadened because they are formed in the hot chromosphere, so they are blended at the majority of epochs. \label{fig:HIRESlines}} 
  \end{figure*}

In Table~\ref{tab:ewha}, we list the equivalent widths of H$\alpha$ emission for those epochs where the lines from the two stars were fully resolved (i.e., within $\Delta \phi < 0.1$ of orbital quadrature). The equivalent widths are measured with respect to the continuum of the full composite spectrum, but individual stellar values can be determined from the flux ratio of the spectra (which is nearly constant across the entire wavelength range of the HIRES spectra). We also show narrow wavelength ranges around H$\alpha$ in each spectrum in Figure~\ref{fig:HIRESlines}. As can be seen, while the H$\alpha$ emission of the primary is clearly evident, the emission from the secondary is weaker and can only be measured in aggregate by measuring the excess flux over continuum within $\pm 1$ \AA\, of the expected line center.

To estimate $v \sin(i)$ from $\sigma_v$, we artificially broadened each of our template spectra using the IDL task {\em lsf\_rotate} \citep{Gray:1992vl,Hubeny:2011ad} using a range of rotational velocities, and then computed each artificial spectrum's broadening function using the corresponding original template. This process yielded a relation between $v \sin(i)$ and $\sigma_v$ that is appropriate for any spectrum observed with Keck/HIRES at this spectral resolution, and can be applied to the $\sigma_v$ values that emerge from the Gaussian function that we fit to each component's broadening function peak in our analysis. Using the 10 spectra with the lowest inter-order scatter of $\sigma_v$, we find that the mean and standard error of the line broadening for each star are $\sigma_{v,p} = 7.78 \pm 0.02$ km/s and $\sigma_{v,s} = 7.55 \pm 0.05$ km/s, which correspond to $v_p \sin(i) = 2.6 \pm 0.6$ km/s and $v_s \sin(i) < 2.0$ km/s, respectively.

\begin{figure*}
 \includegraphics[scale=0.46,trim=0cm 0cm 0cm 0cm]{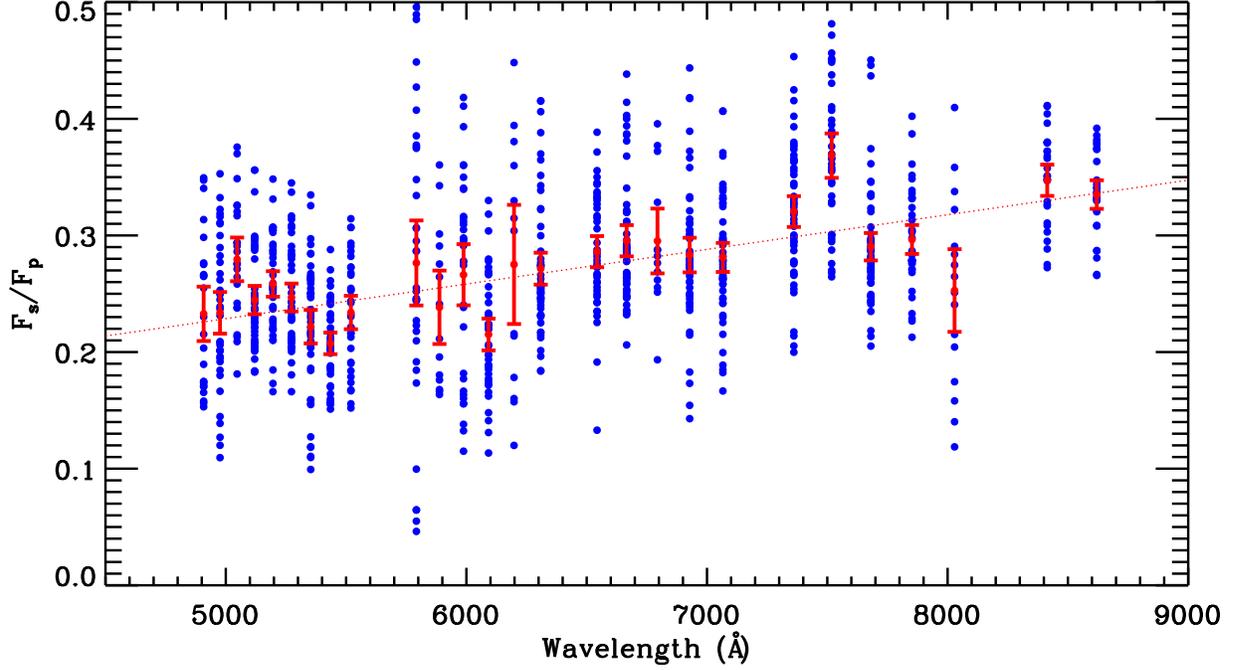}
 \caption{The flux ratio between the binary components as a function of wavelength, as inferred from the ratio of areas under the broadening function peaks. Blue points show individual measurements for each order of each spectrum, with respect to each of the three RV standards. Red points with error bars show the mean and standard error for each order after sigma clipping outliers with a 2$\sigma$ clip. There is a clear linear trend for the secondary to contribute a larger fraction of the total flux at longer wavelengths: $\frac{F_s}{F_p} = [2.9709 \times 10^{-5} \AA^{-1}] \lambda + 0.080$. This trend indicates that the secondary is indeed cooler than the primary. \label{fig:FRATplot}}
 \end{figure*}

In Figure~\ref{fig:FRATplot}, we show the flux ratio measured for the two stars, as computed from the broadening functions for each order of each spectrum, as well as the mean and standard error for each order after applying a 2$\sigma$ clip. The flux ratio should depend on wavelength for two stars with unequal temperatures, and we find that there is a positive linear correlation with slope $a = 2.97 \times 10^{-5} $\AA$^{-1}$. As we discuss below, the points constituting this wavelength-dependent relation provide an important constraint in deconvolving the combined-light spectra taken with SNIFS and SpeX.

\subsection{Intermediate-Resolution Spectroscopy} \label{sec:obs:lores}

We obtained an optical spectrum of \ptfeb on 2016 April 3 with the Supernova Integral Field Spectrograph (SNIFS; \citealt{Aldering:2002rt,Lantz:2004yq}) on the University of Hawai'i 2.2m telescope on Maunakea. SNIFS covers 3200--9700 \AA\, simultaneously with a resolution of $R \sim 700$ and $R\sim 1000$ in the blue (3200-5200 \AA) and red (5100--9700 \AA) channels, respectively. We observed with a total integration time of $3 \times 1200 = 3600$ s, yielding S/N$=100$ per resolving element at $\lambda = 6500$ \AA. We also observed spectrophotometric standards for flux calibration and obtained a ThAr arc before or after each observation for wavelength calibration. Bias subtraction, flatfielding, dark correction, cosmic ray rejection, construction of data cubes, and extraction of the final spectrum were performed as described in detail by \citet{Aldering:2002rt}. The flux calibration is derived from the combination of the spectrophotometric standards and a model of the atmospheric absorption above Maunakea as described by \citet{Mann:2015ys}.

We obtained a NIR spectrum of \ptfeb on 2016 April 5 with the SpeX spectrograph \citep{Rayner:2003vn} on the NASA Infrared Telescope Facility (IRTF) on Maunakea. SpeX observations were taken in the short cross-dispersed (SXD) mode using the 0.3 $\times$ 15\arcsec\, slit, yielding simultaneous coverage from 0.8--2.4 $\mu$m at a resolution of $R \sim 2000$. The target was observed in an AB nod pattern to allow for sky subtraction. We took a total of 34 exposures totaling $4080$ s, yielding S/N$=100$ per resolving element at $\lambda = 2.2$ $\mu$m. Spectra were extracted using the SpeXTool package \citep{Cushing:2004mz} which performs flatfielding, wavelength calibration, sky subtraction, and extraction of the final spectrum. Exposures were combined using the IDL routine {\em xcombspec}. Telluric lines were corrected using a spectrum of the A-type star HD 68703, which was observed immediately before the target with a difference of $<0.1$ airmass, and the correction was computed and applied using the {\em xtellcor} package \citep{Vacca:2003rt}.

Following the method outlined by \citet{Mann:2015ys}, we combined and absolutely flux-calibrated the optical and NIR spectra using published photometry (Section~\ref{sec:obs:lores}) with the filter profiles and zero points provided by \citet{Fukugita:1996ta} and \citet{Mann:2015yz}.

\section{Analysis} \label{sec:analysis}

\subsection{Atmospheric Properties and Radius Ratio from Spectra} \label{sec:analysis:prior}

We initially analyzed the system as a single, unresolved object. Following \citet{Mann:2015ys}, we combined the optical and NIR spectrum (Section~\ref{sec:obs:lores}), which we simultaneously flux calibrated using available photometric measurements (Section~\ref{sec:obs:arcphot}) and the appropriate zero-points and filter profiles \citep{Cohen:2003yu,Mann:2015yz}. We filled in missing regions of the spectrum and areas of high telluric contamination with the best-fit BT-SETTL atmospheric model  \citep{Allard:2011qf}. Once combined and calibrated, we dereddened the spectrum by $E(B-V) = 0.027 \pm 0.004$ mag \citep{Taylor:2006vq}, and then integrated over wavelength to compute the bolometric flux. Accounting for errors in the flux calibration, photometry, photometric zero-points, and reddening, we derived a final bolometric flux of $F_{bol} = (1.75 \pm 0.06) \times 10^{-11}$ erg cm$^{-2}$ s$^{-1}$. To compute the bolometric luminosity, we adopted the distance $d = 182 \pm 6$ pc \citep{van-Leeuwen:2009zr} and found $L_{bol} = 0.0180 \pm 0.0010 L_{\odot}$.

The luminosities and temperatures of the individual stars are subject to a strong joint constraint from the unresolved magnitudes and spectra of the \ptfeb system, when combined with Praesepe's known distance and reddening. However, the colors and molecular absorption bands of M dwarfs vary smoothly and monotonically with temperature, so the same unresolved features are degenerately consistent with a range of plausible temperature and luminosity combinations. To determine non-degenerate temperatures and luminosities for each star, we therefore must also use the wavelength-dependent flux ratio inferred from our Keck/HIRES observations (Section~\ref{sec:obs:hires}; Figure~\ref{fig:FRATplot}). The same analysis also provides a useful prior for our light curve analysis. While eclipse light curves strongly constrain the sum of the stellar radii, they offer a weaker constraint on the ratio of the radii; a very similar lightcurve results from making one star larger and the other smaller, and then optimizing the inclination to match.

We have combined all of these data in a simultaneous fit against a library of empirical, flux-calibrated spectra. We adopted these library spectra from the large sample of nearby M dwarfs considered by \citet{Mann:2015ys}. These stars have high-quality measurements of their distances $d$ (from parallax), metallicities [Fe/H] (from spectra; \citealt{Mann:2013vn}), bolometric fluxes $F_{bol}$ (from spectra and panchromatic broadband photometry), and effective temperatures $T_{eff}$ (from colors and spectra, using a relation bootstrapped from stars with interferometric radius measurements;\citealt{Boyajian:2012il,Mann:2013fj}). Using the Stefan-Boltzmann law and the known values of $d$, $F_{bol}$, and $T_{eff}$, we computed the radius of each library star. We then combined it with the absolute flux-calibrated spectra to compute emergent spectral flux density or surface brightness ($S_{\lambda}$, in erg s$^{-1}$ cm$^{-2}$ \AA$^{-1}$) of the star, as well as the emergent spectral flux densities when averaged across each order of our HIRES spectra and when convolved with the \Kepler and PTF bandpasses ($S_{Kep}$ and $S_{PTF}$).

After constructing this library, we then combined all possible pairs of templates with metallicities consistent with Praesepe ([Fe/H]=0.14$\pm$0.04; \citealt{Taylor:2006vq}) and compared them to the absolutely calibrated unresolved spectrum (Section~\ref{sec:obs:lores}) and spectrally resolved HIRES flux ratios (Section~\ref{sec:obs:hires}) of the \ptfeb A+B system. For each system, our analysis explored the range of allowed total flux ratios (and hence radius and surface brightness ratios) that was consistent with the Keck/HIRES and SNIFS+SpeX results. For each possible combination, we solved for the component stellar radii that would best reproduce the absolute and relative flux measurements of \ptfeb and adjusted the total brightness of the template spectra as appropriate. From the scaled spectra we computed the radius ratio ($\frac{R_s}{R_p}$), \Kepler bandpass flux ratio ($\frac{S_{s,Kep}}{S_{p,Kep}}$), PTF bandpass flux ratio ($\frac{S_{s,PTF}}{S_{p,PTF}}$), and the $\chi^2$ of the fit as dependent variables. 

The $\chi^2$ goodness-of-fit statistic is poorly defined for spectra that have a very large number of wavelength channels and errors that are dominated by the covariance between channels. These covariances can be integrated into the analysis using tools such as {\em Starfish} \citep{Czekala:2015mz}, but the runtime for a large spectral library would be infeasibly long. To avoid having the fit dominated by the spatially unresolved spectra, we instead weighted the final $\chi^2$ contributions of the unresolved spectrum to equal twice the combined contributions of the HIRES flux ratio (with 27 degrees of freedom), and rescaled all of the $\chi^2$ values so that the best-fit value would have $\chi_{\nu}^2 = 1$. Adjusting the weighting by a factor of 5 did not change the results in a significant way.

The result of this analysis is a posterior distribution for the radius ratio $\frac{R_s}{R_p}$, the \Kepler bandpass ($K_p$) surface brightness ratio $\frac{S_{s,Kep}}{S_{p,Kep}}$, and the PTF bandpass (Mould $R$) surface brightness ratio $\frac{S_{s,PTF}}{S_{p,PTF}}$. There are several confounding variables (such as metallicity, stellar age, and spot coverage), and even small errors in the flux calibration can lead to significant spectral mismatch across many channels, so there is not a smooth $\chi^2$ hypersurface within this three-dimensional space. Adjacent points differ significantly in $\chi^2$. We instead constructed the joint posterior using the 9276 template pairs with a goodness-of-fit $\chi_{\nu}^2 < 4$ (chosen to yield visually acceptable matches to the spectra), taking the density of fit results in the three-dimensional space as a measure of the posterior probability density. To avoid establishing a prior that goes to zero outside of the distribution of points, but instead declines smoothly away from this locus, we defined the density at a given location in parameter space by convolving each discrete point with a 3D gaussian blurring function, with $\sigma = 0.05$ on all axes. We have verified that the shape of this posterior does not change significantly for $\chi^2$ cuts or different values of $\sigma$, even extending to much poorer fits ($\chi_{\nu}^2 \sim 10$) where the spectral mismatch is visually obvious. In Figure~\ref{fig:RFFpost}, we show the distribution of 9276 points that defines our posterior distribution for $\frac{R_s}{R_p}$, $\frac{S_{s,Kep}}{S_{p,Kep}}$, and $\frac{S_{s,PTF}}{S_{p,PTF}}$.

In addition to computing a spectroscopic prior for our MCMC orbit analysis, we also used the same template library and fitting scheme to estimate posteriors for the best-fit temperatures, spectral types, and bolometric fluxes for the components of \ptfebb. To take advantage of the strong constraints on the surface brightness ratio that emerge from the light curves (Section~\ref{sec:analysis:mcmc}), we computed our MCMC analysis of the light curve without using the prior on the surface brightness ratios and radius ratio (to avoid double-weighting the flux ratio constraints from the spectroscopic observations) and then used the resulting posteriors on $\frac{R_s}{R_p}$, $\frac{S_{s,Kep}}{S_{p,Kep}}$, and $\frac{S_{s,PTF}}{S_{p,PTF}}$ as input priors for the analysis described above. We adopted the resulting set of all template pairs with $\chi_{\nu}^2 < 4$ as a posterior distribution for the individual component temperatures, spectral types, bolometric fluxes, and bolometric luminosities. We find that the two template stars that produce the lowest overall $\chi^2$ are HD 18143 C ($T_{eff} = 3227$ K; [Fe/H]$= +0.28 \pm 0.03$) and GJ 3668 ($T_{eff} = 3109$ K; [Fe/H]$= -0.07 \pm 0.08$).

\subsection{MCMC Fitting for Orbital and Stellar Parameters} \label{sec:analysis:mcmc}

We have fit the system properties using an updated version of the Markov-Chain Monte Carlo (MCMC) procedure that we described in more detail in our analysis of the low-mass eclipsing binary UScoCTIO 5 \citep{Kraus:2015mz}. To briefly summarize, our pipeline simultaneously fits the RV curve and all available light curves with a model consisting of six orbital elements ($T_0$, $P$, $a$, $e$, $\omega$, and $i$), the mass ratio of the system $q = M_s/M_p$, the systemic radial velocity $\gamma$, the sum of the stellar radii $R_{tot}=R_p+R_s$, the ratio of the stellar radii $r=R_s/R_p$, and the ratios of stellar fluxes through the \Kepler $K_p$ bandpass and PTF $R$ bandpass. 

We fit the RV curves with analytically determined radial velocities at each epoch; none of our spectra were taken during eclipse, so we do not need to include the Rossiter-McLaughlin effect. We fit the light curves with an analytic formalism based on the work of \citet{Mandel:2002ai}, with modification to allow for luminous occulting bodies. The K2 exposure time (29.4 minutes) is long compared to the typical change in brightness during eclipses, so we modeled 27 sub-exposures of duration 65.4 seconds, and then summed those fluxes for comparison to the observations. To model limb darkening, we use a quadratic relation with the coefficients prescribed for a star of appropriate $T_{\rm eff}$ and $\log(g)$ by \citet{Claret:2012tg}: $\gamma_{1,P,R} = 0.6171$, $\gamma_{2,P,R} = 0.3327$, $\gamma_{1,S,R} = 0.5436$, $\gamma_{2,S,R} = 0.2532$, $\gamma_{1,P,Kp} = 0.4930$, $\gamma_{2,P,Kp} = 0.4298$, $\gamma_{1,S,Kp} = 0.4488$, and $\gamma_{2,S,Kp} = 0.2818$.

Our algorithm was chosen due to its fast runtime, which allows for efficient exploration of our high-dimensional model by our MCMC, but this design choice also comes with necessary caveats. We do not include several physical effects that are modeled in more sophisticated code (e.g., \citealt{Wilson:1971ig}), such as reflected light and ellipsoidal variation. However, those effects are negligible for main-sequence stars in well-detached systems. More significantly, our code also does not include any model for starspots. As can be seen in Figure~\ref{fig:k2curve}, the photometric amplitude of the system outside of eclipse ($\pm$3\%) indicates the presence of large and complicated spot complexes. Occultation of those spots will introduce high-frequency noise in the eclipse light curves. Traditionally, these spots are fit with a spot model that is consistent with the out-of-eclipse variations, implicitly rectifying the variations in system flux. However, spatially unresolved photometry does not contain sufficient information to reconstruct a unique distribution of starspots across the stellar surfaces, so the variations are typically modeled with 1--2 very large spots. These incorrect spot models will degrade the precision of the eclipse fit by simultaneously not encompassing the fine details of the spot structure (which can not be fit from the variations in total system flux) and forcing the fit to account for a spot model that is not correct. As we discussed in \citet{Kraus:2011lr}, uncorrected spots result in radius variations of $\pm$2\%; we explore this possibility in further detail in Section~\ref{sec:results:spots}.

We have modified several aspects of our pipeline since our analysis of UScoCTIO 5 in \citet{Kraus:2015mz}:

\begin{itemize}

\item Multiple filters: Since we have multi-band photometry for \ptfeb ($R_{PTF}$ and $K_p$, albeit not simultaneously), we now fit for a surface brightness ratio in each of these bandpasses.

\item Spectroscopic flux ratios: We previously used the optical flux ratio inferred from the broadening functions of each star in Keck/HIRES spectra (Section~\ref{sec:obs:hires}; Figure~\ref{fig:FRATplot}) as a direct constraint on the radius ratio and the $K_p$ surface brightness ratio, $\frac{F_s}{F_p} = \frac{S_{s,Kep}}{S_{p,Kep}} \times (\frac{R_s}{R_p})^2$. However, this choice did not fully exploit the measurable wavelength dependence of the HIRES flux ratio. We now incorporate the Keck measurements into the analysis of the component temperatures and bolometric fluxes (Section~\ref{sec:analysis:prior}), and use the posterior joint constraints on $\frac{R_s}{R_p}$, $\frac{S_{s,Kep}}{S_{p,Kep}}$, and $\frac{S_{s,PTF}}{S_{p,PTF}}$ as priors for our MCMC fit of the RV and light curves.

\item Fitting $T_{P}$ instead of $T_0$: For eclipsing systems that are nearly circular, the combination of the longitude of periastron $\omega$ and the time of periastron $T_0$ are highly degenerate and poorly constrained. A Gibbs sampler that separately explores these parameters will mix very slowly due to this degeneracy. We therefore have modified our code to fit the time of primary eclipse $T_{P}$ (which is very well constrained by the eclipse photometry) and $\omega$, and then to compute $T_0$ as a dependent quantity. The net result is equivalent to an MCMC that explores on a linear combination of $T_0$ and $\omega$ (but without the need to calculate this linearization explicitly for each system) or that explores on $\sqrt{e}\cos(\omega)$ and $\sqrt{e}\sin(\omega)$\citep{Eastman:2013yg}.

\end{itemize}

We use a uniform prior for all variables. The eccentricity is not bounded at $e=0$; if a jump reduces the eccentricity to $e < 0$, then the eccentricity is made positive and $\omega$ is rotated by 90\degr. The mass ratio is not bounded at $q=1$, allowing for the star labeled as the secondary to become more massive. If a jump would increase the inclination to $i > 90\degr$, then the inclination is set to $180\degr - i$.

We executed the MCMC using a Metropolis-Hastings sampler to walk through parameter space, selecting jump sizes and establishing initial burn-in using test chains from a range of starting parameter states. For the final parameters we computed 20 simultaneous chains for a total length of $1.1 \times 10^5$ steps per chain, omitting the first $10^4$ steps of each chain to allow for random dispersal from the (common) initial starting point. As a result, our distributions have $2 \times 10^6$ distinct samples from which the posteriors on each parameter are constructed. We verified that the individual chains yield mean values that agree to within much less than the reported $1\sigma$ uncertainties, indicating that they are well-mixed. We also calculated other parameters of interest ($M_p$, $M_s$, $M_{tot}=M_p+M_s$, $R_p$, $R_s$) from the fit parameters at each step in the chain, yielding similar posterior distributions. Finally, we explored the robustness of our results by fitting many subsets of the data (Section~\ref{sec:results:subsets}), where for each subset we computed 20 simultaneous chains for a total length of $2 \times 10^4$ steps per chain, omitting the first $10^4$ steps for burn-in and dispersion from the initial starting point.

As we discuss further in Appendix A, we have validated our pipeline by analyzing extant data for the well-studied system GU Boo (LMR05), showing that our very different analysis methods match previous radius measurements to within 2--3\%. A similar result was found by \citet{Windmiller:2010rs} using the ELC software \citep{Orosz:2000fk} and additional data on the GU Boo system.

\section{Results} \label{sec:results}

\subsection{System Properties} \label{sec:results:sysprops}

\ptfeb is one of the few low-mass ($M_p \la 0.7$~$M_{\odot}$) eclipsing binaries to be found in an open cluster (see, e.g., \citealt{David:2015aa,David:2016aa}), and therefore it poses a test of main sequence stellar models for which the metallicity ([Fe/H]$ = 0.14 \pm 0.04$; \citealt{Taylor:2006vq}), distance ($d = 182 \pm 6$ pc; \citealt{van-Leeuwen:2009zr}), and age ($\tau \sim 600$--800 Myr; \citealt{Delorme:2011qb,Brandt:2015fv}) are not confounding free parameters. Furthermore, the long orbital period and lack of tidal locking suggest that the properties of the two stars are broadly representative of typical young ZAMS stars, in contrast to the short period, tidally locked rapid rotators that comprise most of the EB sample studied to date.

We summarize our best-fit properties of \ptfeb and its components in Table~\ref{tab:partable}, and in Figures~\ref{fig:rvfit}, \ref{fig:lcfit}, and \ref{fig:ptffit}, we show the observed RVs and \Ktwo/PTF photometry, the best-fit model RV curve and light curves, and the residuals between the observations and the data. We find that \ptfeb consists of stars with unequal masses ($M_p = 0.395 M_{\odot}$, $M_s = 0.210 M_{\odot}$) and radii ($R_p = 0.36 R_{\odot}$, $R_s = 0.27 R_{\odot}$). The fractional uncertainties on the individual masses are $\la$1\% due to the dense phase coverage and excellent instrumental stability of the Keck/HIRES data. The fractional uncertainties on the primary and secondary radii are $\sim$2\% and $\sim$4\% respectively. We discuss possible systematic errors in the radius measurements in Sections 4.2, 4.3, and 4.4.

In Figure~\ref{fig:posteriors1d}, we show the marginalized one-dimensional posterior distributions of each parameter, as well as the median and central 68\% credible interval. Most posteriors are distributed symmetrically about the median, and therefore the median and central interval are good representations of the most likely value. The clearest exception to this case is the eccentricity distribution, for which the mode of the distribution ($e \sim 0.0013$) is located just outside the lower edge of the central 68\% credible interval. However, the difference between the median and mode does not change any astrophysically useful results. The other clearly asymmetric distributions are those of $\omega$ and $T_0$, which are tightly correlated (Section~\ref{sec:analysis:mcmc}), but again the detailed values do not impact our conclusions.

In Figure~\ref{fig:corner}, we show a triangle plot of the six astrophysically important parameters that are most likely to be degenerate with each other: $e$, $i$, $R_p + R_s$, $\frac{R_s}{R_p}$, $\frac{S_s}{S_p}_{Kep}$, and $\frac{S_s}{S_p}_{PTF}$. The only apparent degeneracies among parameters are those that are well known for eclipse binary analyses. Most significantly, the radius ratio $\frac{R_s}{R_p}$ is tightly correlated with the inclination, since the fractional occulted area (and hence the depth and duration of the transit) only changes very gradually while changing the impact parameter and the relative stellar radii. However, changing the inclination does change the total sum of the radii that is needed to preserve the eclipse morphology, so there is also a looser correlation between $R_{tot}$ and both $\frac{R_s}{R_p}$ and $i$.

The stellar rotation periods can be inferred from both the light curve (for the primary star) and the high-resolution spectra (for both stars, and assuming spin-orbit alignment). The light curve is dominated by flux from the primary star, contributing $\sim$75\% of the flux in the red optical (Figure 6). If the observed sinusoidal variations (with full amplitude 6\%) were caused by the secondary star, then its individual total amplitude of variation would be 26\%; studies of rotational variability across the full sample by \citet{Douglas:2017aa} and \citet{Rebull:2017aa} found that the maximum amplitude seen for 0.2 $M_{\odot}$ stars was only 10\%. A similar upper envelope has been seen for periodic field stars observed by Kepler by \citet{Harrison:2012fk}. We therefore conclude that the photometric variations outside of eclipse (Section~\ref{sec:obs:k2} and Figure~\ref{fig:k2curve}) show that $P_{rot,P} = 7.46$ d. 

Given our measured radius, the corresponding rotational broadening of our spectra would be $v_p \sin(i) \sim 2.5$ km/s, which is consistent with the measured rotational broadening of $v_p \sin(i) = 2.6 \pm 0.6$ km/s. The rotational signature of the secondary star is not evident in our light curves, but our measured radius and the upper limit on $v \sin(i)$ from our Keck/HIRES spectra ($v_s \sin(i) < 2.0$ km/s) imply a rotational period of $P_{rot,S} \ga 6$ d. Our measurement would be consistent with tidal locking of the secondary, but we can not verify whether this has occurred. The rotational period of the 0.4 $M_{odot}$ primary makes it a normal object on the $P_{rot}$-$M$ relation for Praesepe \citep{Douglas:2017aa,Rebull:2017aa}, but it is noticeably faster than the ensemble of field 0.4 $M_{\odot}$ stars (e.g., \citealt{Harrison:2012fk,McQuillan:2013mz,Newton:2016aa}), suggesting that it is a suitable representative of a young ZAMS star. There is only a lower limit on the rotational period of the 0.2 $M_{\odot}$ secondary; at that limit, it would sit on the slow edge of the Praesepe sequence, but would not be unusually slow.

As we show in Figures~\ref{fig:posteriors1d} and \ref{fig:corner}, the best-fit eccentricity for the system is very small, but is not zero. This result emerges directly from the \Ktwo light curve. Figure~\ref{fig:lcfit} demonstrates that the secondary eclipse is $\sim$$10^{-3} P_{orb}$ ($\sim$20 minutes) earlier than the halfway point between primary eclipses. This small eccentricity can not be detected in the RV curve or PTF light curve alone, and the longitude of periastron is not tightly constrained ($\omega = 45 \pm 25 \degr$). An azimuthally asymmetric brightness distribution on one of the stars (due to spots) could cause an apparent shift in the measured eclipse midpoint. The primary star is not tidally locked though, so an azimuthal asymmetry on that star would cause stochastic variations in eclipse timing, not a constant offset. We can not rule out this hypothesis for the secondary star, since the lower limit on its rotational period would be consistent with tidal locking.

The surface temperatures of the stars can be inferred in two complementary ways. Our spectroscopic analysis (Section~\ref{sec:analysis:prior}) yields best-fit temperatures of $T_{eff,P} = 3260 \pm 30$ K and $T_{eff,S} = 3120 \pm 50$ K, in both cases subject to a 60 K (0.5 subclass) systematic uncertainty from the definition of the underlying grid. Our measurements of the bolometric luminosity and the radius of each star give geometric measurements that are independent of any spectral classification system, albeit with a large uncertainty for the secondary star, yielding $T_{eff,P} = 3290 \pm 70$ K and $T_{eff,S} = 2970 \pm 230$ K. We therefore find good agreement for the primary to have $T_{eff} \sim$3250--3300 K, while the secondary is most likely $\sim$150 K cooler than the primary star.

Our measurement of the systemic velocity of \ptfeb ($\gamma = 34.00 \pm 0.15$ km/s) is consistent with the typical range seen for Praesepe members ($v_{rad} \sim 33$--34 km/s; \citealt{Mermilliod:1999yf}). The proper motion was already known to closely agree; \citet{Kraus:2007mz} found $\mu = (-37.5, -14.1) \pm 2.7$ mas/yr, which agrees with the mean HIPPARCOS value to within $< 1 \sigma$. Given the HIPPARCOS distance, we find that the corresponding space velocity for \ptfeb is $v_{UVW} = (33.8 \pm 1.7, -8.5 \pm 2.2, -2.5 \pm 2.1)$ km/s.

Finally, we note that while this paper was under review, \ptfeb was also reported as an eclipsing binary by \citet{Rebull:2017aa}, \citet{Douglas:2017aa}, and \citet{Gillen:2017aa}. The latter group analyzed 8 RV measurements and the K2 light curve to compute system parameters. They report masses that broadly agree with our results ($\sim 2 \sigma$ smaller, based on mutual uncertainties), as well as a similar primary star radius. However, they report a significantly smaller secondary star radius ($R_{s} = 0.226 R_{\odot}$ versus $R_{s} = 0.272 R_{\odot}$, a $3.4 \sigma$ discrepancy); this measurement appears to result from a substantially smaller radius ratio estimate that emerges from their spectroscopic prior. These differences further emphasize the need to understand systematic differences that emerge from different analysis pipelines (Sections 4.2, 4.3, 4.4, and Appendix A).

\subsection{The Robustness of Eclipsing Binary Fits} \label{sec:results:subsets}

To test the robustness of our results, we repeated our analysis using only subsets of the data. The radial velocity measurements are always required, since they yield a unique measurement of the component masses. However, the geometric and surface properties are overconstrained by the combination of the \Ktwo light curve, the PTF light curve, and the spectroscopic prior. We therefore can omit subsets of these data while still finding well-bounded posteriors, and hence can determine both the importance of each data source and whether all data sources are indicating consistent system properties. 

We summarize the variation in these properties when fitting different subsets of the full dataset in Table~\ref{tab:comptable}. We specifically list the eccentricity and inclination of the orbit, the sum and ratio of the stellar radii, the individual stellar radii, and the ratio of the surface brightnesses in the \Kepler and PTF bandpasses. We have executed separate MCMC runs by omitting either all \Ktwo primary eclipses, all \Ktwo secondary eclipses, all PTF data, or the spectroscopic prior, as well as all combinations thereof that lead to bounded posterior distributions.

The system parameters are surprisingly robust when omitting data sources, changing by $\la 3 \sigma$ and generally with only modest increase in the uncertainty. There is only a modest impact on the inferred radii. When omitting one data source, the radius fits for the primary star span 0.353--0.377 $R_{\odot}$ ($\pm 3.3\%$), suggesting that the measurement is robust and all data ultimately point to similar values. The equivalent radius fits for the secondary star span 0.260--0.285 $R_{\odot}$ ($\pm 4.5\%$), again consistent to within the uncertainties. It is also relevant to consider omission of multiple data sources; while \ptfeb is exceedingly well characterized, most systems discovered in \Ktwo or other programs will lack such an abundance of data. When omitting any two data sources, the radius fits for the primary and secondary star span 0.338--0.387 $R_{\odot}$ ($\pm 6.8\%$) and 0.246--0.308 $R_{\odot}$ ($\pm 11 \%$) respectively. In all cases, the fit parameters inferred from the full dataset are centrally located within the range of parameters inferred for the subsets. As we discuss further in the Section~\ref{sec:results:spots}, this robustness has strong implications for the impact of spots on our test of the stellar mass-radius relation.

The impact of removing (over)constraints can be seen more clearly in the remaining fit parameters. The sum of the radii, which is strongly constrained by the total duration of the eclipses, remains nearly constant (spanning 0.615--0.647 $R_{\odot}$) and well-constrained (with error $<$3\%) in all cases. Even fitting only the PTF light curve still yields the radius sum with 3\% uncertainty. In contrast, the radius ratio becomes poorly constrained in several subsets and is effectively unconstrained (allowing a ratio above unity at $\la 3 \sigma$) when omitting both the prior and another dataset. The degeneracy can be avoided for totally eclipsing systems with flat eclipse minima, where the flux ratio securely measures the radius ratio, but few systems meet this geometric requirement. We therefore suggest that population analyses of known low-mass EBs from the literature (which often are not observed so comprehensively) should compare observed and theoretical radius sums, rather than attempting to consider the two individual stellar radii for each system.

\subsection{The Influence of Spots on Radius Measurements} \label{sec:results:spots}

M dwarfs commonly host substantial starspots, particularly at young ages (e.g., \citealt{Cody:2014uq}) or when tidally locked into fast rotation \citep{Lopez-Morales:2007kz}. These starspots complicate the analysis of eclipsing binary light curves. The most obvious impact is that variations in the total spot coverage across both stars will change the out-of-eclipse brightness as a function of rotational phase, requiring rectification in order to properly measure the decrement in brightness specifically due to eclipses. However, the more severe impact for determining detailed stellar properties occurs during eclipse. Changes in relative spot coverage on the occulted area (on the eclipsed star) and the unocculted areas (on part of the eclipsed star and all of the eclipsing star) will change the overall amplitude of the eclipse, and occultations of individual spots on the background star will introduce high-frequency noise into the eclipse light curve. Most eclipsing binaries have short orbital periods and are tidally locked \citep{Zahn:1977ni}, so the same hemisphere of each star is always visible during eclipse. The effect is therefore identical on all timescales shorter than the spot evolution timescale (i.e., years; \citealt{Morales:2009rb,Windmiller:2010rs}), making it difficult to measure the impact or to average out the effect with more data.

\ptfeb offers a unique opportunity to conduct repeatable tests of the impact of spots on stellar radius determinations. The ratio between the orbital period of \ptfeb (6.016 d) and the rotational period of the primary star (7.46 d) is very close to 4:5, so each primary eclipse occults almost exactly the same range of longitudes (on the primary star) as was occulted 30 days earlier or later. Among the 12 primary eclipses that occurred during the continuous 80 day \Ktwo observation, there are two occultation configurations that recur three times (eclipses 1+6+11 and 2+7+12) and three more that recur twice (eclipses 3+8, 4+9, and 5+10). However, primary eclipse \#10 occurred shortly after a flare; we omitted it from our earlier analysis, and while we include it in our tests, we similarly do not factor the results into our conclusions.

To test for potential systematic errors from the unknown spot configuration, we first reran our MCMC 12 times, each time masking all but one primary eclipse to create a subset of the data. We found that the posterior distribution for the system properties did not substantially change, but this result is predictable. As we discussed in Section~\ref{sec:results:subsets}, our extensive dataset overconstrains the system parameters. If accurate and precise radii can be inferred without any primary eclipses, it naturally follows that any single measurement does not substantially impact the fit, especially since there are only 6 photometric measurements during each eclipse. However, most systems are unlikely to be characterized this well. We therefore have repeated the test with only the \Ktwo data and spectroscopic prior, omitting the PTF dataset and using a single \Ktwo primary eclipse at a time. 

In Figure~\ref{fig:radfig}, we show the resulting posterior distributions for the individual stellar radii using our fit to the \Ktwo light curve and spectroscopic prior, the \Ktwo secondary eclipses and prior, and the intermediately constrained cases with each of the 12 \Ktwo primary eclipses, the \Ktwo secondary eclipses, and the prior. In each panel, we outline the 68\% credible intervals for the joint posterior on the primary and secondary radii in that case. We find that using fewer datapoints for the primary eclipse yields larger uncertainties on the stellar radii, as we would expect in a regime where Poisson errors dominate. However, all of the test fits yield similar values to within 1$\sigma$, and most are within 1$\sigma$ of the value when fitting all of them. This result suggests that even though the primary star is heavily spotted (leading to $\pm 5\%$ total flux variations), changes in the detailed spot configuration do not significantly impact our results. We find an RMS scatter of 0.7\% for $R_p + R_s$, 1.8\% for $\frac{R_s}{R_p}$, 0.4\% for $R_p$, and 1.7\% for $R_s$.

Visual inspection suggests that eclipses of the same hemisphere (denoted by curves with the same color) might more closely resemble each other than the ensemble as a whole, but if so then only very modestly. If the scatter were purely Poisson, but with no systematic effect resulting from sampling five specific spot patterns, then averaging all eclipses of the same hemisphere should reduce the scatter to $\sqrt{\frac{5}{11}}$ or 67\% of these values. We find that the reduction is indeed by approximately this amount; the RMS scatter across the five values is reduced to 0.5\%, 1.0\%, 0.4\%, and 1.0\%, respectively, corresponding to 69\% of the original scatter. Our results suggest that any impact from the detailed spot configuration is less than the radius uncertainties resulting from our analysis ($\sim$2\%).

We note that these measurements are unavoidably noisy due to having only 11 eclipses that are distributed between five different spot configuration states, so the exact impact remains uncertain. Given the large uncertainty of the radius ratio and the robustness of the radius sum, unequal-radius systems will see disproportionately greater impact on the secondary, as we find here. However, we broadly conclude that using optical light curves that sample only a single spot configuration results in a characteristic noise floor of no more than 1--2\%, consistent with the results of our earlier simulations \citep{Kraus:2011lr}. Any further reduction of this noise floor would require either observations at longer wavelengths (where spots have lower contrast), sampling multiple spot configurations (by observing systems that are not tidally locked or observing locked systems for longer than the spot evolution timescale), or using outside constraints like multi-color photometry (such as our use of PTF data).

\subsection{Systematic Uncertainties in Radii from Limb Darkening Models}

The morphology of an eclipse light curve and the inferred stellar radii also depend on the stellar limb darkening, since a limb-darkened star would be visually similar to a uniformly illuminated disk of smaller radius. Nearly the same eclipse can result from a degenerate combination of possible limb darkening laws, stellar radii, and orbital inclinations, so assumptions about the limb darkening law will be reflected directly in the resulting radius measurements. Due to the paucity of spatially resolved images of stars, limb darkening laws are typically derived from radiative transfer through model atmospheres, but different treatments yield prescriptions that differ by $\sim$0.1 even just in the linear term (e.g., \citealt{Claret:2011aa} vs \citealt{Sing:2010aa}). Some light curve analyses attempt to bypass the theoretical uncertainties by directly fitting for the limb darkening coefficients, but these empirical measurements do not yield consistent results between different stars and can sometimes even imply naively unphysical results such as limb brightening. The empirical analyses might themselves be limited by spatial inhomogeneities in the surface brightness of the stars (such as from spots and plages) that can change the de facto limb darkening of the star, especially in the case of spatially coherent spot patterns like polar spots \citep{Morales:2009rb}. 

As with our treatment of spots, we conservatively adopt a uniform theoretical treatment of limb darkening, but then quantify the resulting systematic uncertainty in our radius measurements that could result from potential variations in the limb darkening law. To that end, we have repeated our analysis of the dataset using three other assumptions for the limb darkening. To establish a baseline worst-case scenario, we first removed all limb darkening effects and treated the stars as uniformly illuminated disks. We then repeated our analysis twice more with quadratic limb darkening in place, but with the linear coefficients increased or decreased by 0.1, which establishes a more relevant range of likely radius uncertainties.

In Figure~\ref{fig:LDfig}, we show the resulting posterior distributions for the individual stellar radii using our default limb darkening values and the three modified sets of assumptions. In each case, we outline the 68\% credible interval for the joint posterior on the primary and secondary radii. As expected, we find that the radius depends on the assumed limb darkening law; removing limb darkening entirely results in stars that are $\sim$6\% smaller, while the more realistic variations in the linear term result in $\pm$1--2\% changes in the inferred radius sum and individual radii. We find that for this system, most of the impact from small changes in limb darkening is reflected in the radius of the secondary star. In cases where the limb darkening varies from star to star (such as from spots), then the individual radii and the radius ratio would vary more substantially. However, the broad conclusion is that uncertainties from limb darkening are of order $\sigma_R \sim $1--2\%, and hence they are not any larger than the other systematic effects we discuss.

\section{Confronting Stellar Evolutionary Models} \label{sec:models}

Evolutionary models of M dwarfs are fundamentally a set of relations between independent variables that are fixed for a given star (mass and metallicity), prescriptions for linked features that evolve over time (rotation, magnetic field strength, and spot fraction), and predicted dependent variables (radius, $T_{eff}$, and luminosity). These models are calibrated using observations that probe different combinations of the independent, prescriptive, and dependent variables. Examples include panchromatic spectroscopy to calibrate the luminosity-$T_{eff}$ relation (e.g., \citealt{Mann:2015ys}), long-baseline interferometry to calibrate the luminosity-$T_{eff}$-radius relation (e.g., \citealt{Boyajian:2012il}), and monitoring of visual binaries to calibrate the luminosity-$T_{eff}$-mass relation \citep{Delfosse:2000it,Benedict:2016ul}; all of these tests can be conducted in stellar populations to further probe metallicity and age. Eclipsing binaries pose a unique test of evolutionary models, though, offering the only direct relation between mass and radius, while also being amenable to the measurement of luminosity and $T_{eff}$.

In Figure~\ref{fig:hrdmrd} (left), we plot the locations of the binary components in the HR diagram, along with the coeval isochrones of the BHAC15 \citep{Baraffe:2015qy} and DSEP \citep{Dotter:2008qq} stellar evolutionary models; for the latter, we show tracks at solar metallicity and at the metallicity of Praesepe ([Fe/H]$ = 0.14$). We also denote the model-predicted luminosities and temperatures for each component, given our dynamical mass measurements. We find that both stars are predicted to be hotter than the measured temperatures, and the primary is also predicted to be more luminous than its measured luminosity.

In Figure~\ref{fig:hrdmrd} (right), we plot the locations of the binary components in the mass-radius diagram, also in comparison to the BHAC15 and DSEP models. The radius of the primary star appears to be predicted well by both sets of theoretical models, with a discrepancy within the observational uncertainty. However, both sets of theoretical models predict a significantly smaller radius for the secondary star ($R \sim 0.22$~$R_{\odot}$) than we empirically measure ($R = 0.272 \pm 0.012 R_{\odot}$), disagreeing by $\sim$0.05 $R_{\odot}$ or $\sim$20\%.

The unexpected cool temperature of the primary star appears to be robust, since it dominates the observed flux. Past observations of the (unresolved) system have found a spectral type of M3.4 from the broadband SED \citep{Kraus:2007mz} and M5 from the optical spectrum \citep{West:2011zl}. In comparison, the model-predicted temperature of $T_{eff} \sim 3500$ K corresponds to a spectral type of M1.5 \citep{Casagrande:2008zz} to M2 \citep{Rajpurohit:2013qy}, depending on the temperature scale used. The observed presence of VO bands at 7600--7800 \AA\, make the distinction unambiguous, since they are not present for $\le$M2 stars. Furthermore, the CMD for the rest of Praesepe shows a similar tendency for members to be systematically redder/cooler than models, as demonstrated in Figure 3 of \citet{Mann:2017ys}.

The HR diagram positions are inferred via a method that is almost entirely independent of the radius measurements, so it further bolsters the reliability of our analysis that the discrepancies for the well-characterized primary star are of appropriate sign and magnitude to preserve the Stefan-Boltzmann Law ($L = 4 \pi R^2 \sigma T^4$). The radius is consistent with theoretical predictions, and the discrepancy in predicted temperature ($250 \pm 30$ K or $\sim$7\%) is almost exactly offset by the discrepancy in predicted luminosity ($0.004 \pm 0.001$ $L_{\odot}$ or $\sim$28\%). The combined properties therefore indicate a mutually consistent discrepancy with respect to models. We can not conduct this consistency check for the secondary star because the fractional luminosity uncertainty ($\sim$30\%) is too large for a meaningful comparison.

The origin of these discrepancies remains unclear, and will be difficult to assess with only a single well-characterized system. A number of hypotheses can be deemed unlikely. Models computed at [Fe/H]$ = 0$ and [Fe/H]$ = 0.14$ (appropriate for Praesepe) give virtually identical predictions for stellar parameters, so metallicity mismatch does not seem to be significant. \citet{Stassun:2014lr} found among young eclipsing systems that radii were larger than model predictions in triple systems, which might suggest an origin in dynamically driven binary interactions. We do not have adaptive optics imaging for this system, but the absence of a third broadening function peak suggests that any tertiary companion would need to be substantially fainter ($\ga$2 magnitudes at 8000\AA). Finally, Praesepe is old enough that even the secondary star should have reached the ZAMS much earlier in its lifetime \citep{Dotter:2008qq,Baraffe:2015qy}, so it is also unlikely that the secondary is still undergoing pre-main sequence contraction.

However, while this system is likely on the ZAMS, it is still substantially younger than the field population. Old field stars typically show smaller discrepancies or even agree with models \citep{Torres:2010lr,Kraus:2011lr,Feiden:2014qf}. Young stars are known to be more active than their older counterparts (e.g., \citealt{West:2008wb}), suggesting a possible role for magnetic fields \citep{Chabrier:2007cs,Feiden:2014yu} or starspots \citep{Somers:2015yq} in changing the interior structures of stars or the course of stellar evolution. Magnetism only seems to change the stellar parameters of fully convective main sequence stars if interior magnetic field strengths are far higher than expected ($B > 1$ MG; \citealt{Feiden:2014yu,Feiden:2014qf}). The corresponding surface magnetic field ($B \sim 10$ kG) may be possible in the cores of spots, but exceeds the thermal equipartitian value and is unlikely to represent a global average value (e.g., \citealt{Shulyak:2011aa,Shulyak:2014qv}). The impact of starspots on stellar evolution in mid-M dwarfs should be explored further in theoretical models, though.

Our measurements indicate that the stars in \ptfeb are not unusually active or rapidly rotating compared to other Praesepe members (e.g., Douglas at al. 2014). Given a system flux ratio of $\sim$3:1 at $\lambda = 6500$\AA, the individual H$\alpha$ equivalent widths for each star are $EW[H\alpha]_P \sim -4$\AA\, and $EW[H\alpha]_S \sim -2$\AA; both measurements fall on the mass-activity sequence of Praesepe members (from Figure 5 of \citealt{Douglas:2014nr}), and indeed the secondary falls at the less active edge of the H$\alpha$ distribution for its mass. Similarly, the rotational period of the primary ($P_P = 7.46$ d) falls on the cluster mass-rotation sequence, while the rotation period of the secondary ($P_S \ga 6$ d) is at or beyond the slow edge of the distribution for its mass. We therefore do not find any evidence that the disagreement with models should be limited to this system. Given that the entire Praesepe HR diagram sequence is found to be cooler than models predict, it is likely that the tension with models is a common feature of all low-mass stars at this age.

\section{Summary and Conclusions} \label{sec:summary}

We have discovered and characterized \ptfebb, a $P_{orb} = 6.0$ day eclipsing binary system in the Praesepe open cluster. We find that the system comprises two late-type stars ($SpT_P = M3.5 \pm 0.2$; $SpT_S = M4.3 \pm 0.7$) with precisely measured masses ($M_p = 0.3953 \pm 0.0020$~$M_{\odot}$; $M_s = 0.2098 \pm 0.0014$~$M_{\odot}$) and radii ($R_p = 0.363 \pm 0.008$~$R_{\odot}$; $R_s = 0.272 \pm 0.012$~$R_{\odot}$). Based on tests using subsets of our data, we find that the results are consistent to within 2--3\% even when omitting some datasets. We also find that at least the primary star is not tidally locked to the orbital period, and we take advantage of the 4:5 ratio between $P_{orb}$ and $P_{rot,P}$ to conduct a natural experiment in the variance and repeatability of radius measurements  due to different spot configurations on the occulted star. We demonstrate that the full analysis is resilient to changes in spot variation (changing the inferred radii by $\la$1\%); for data volumes more typical of EB studies, the scatter in inferred radii is still no more than 1--2\%, implying a noise floor in the characterization of most eclipsing systems. We also test different assumptions for limb darkening and show that for a plausible range of parameters, the stellar radii only change by 1--2\%.

Given the masses, neither star meets the predictions of stellar evolutionary models of the appropriate age and metallicity. The primary star has the expected radius, but it is cooler and less luminous than models would predict, while the secondary star has the expected luminosity, but it is cooler and substantially larger (by 20\%) than models would predict. These results broadly match the known discrepancy between theoretical models and Praesepe's empirical cluster sequence in the HR diagram, and the mass and radius measurements now reveal the magnitude of the discrepancies in each quantity. These discrepancies are larger than for old field stars that also are still on the ZAMS, suggesting that age-dependent effects beyond pre-main sequence contraction and nuclear burning (such as rotational spindown and the decay of stellar activity) must be incorporated in stellar evolutionary models. Moreover, the primary and secondary stars span the fully convective boundary, so the different forms of discrepancy might indicate that the existence of a convective-radiative boundary influences the extra physics, as would be expected if tied to generation and field strength of the stellar magnetic fields.

Finally, we validated our pipeline by analyzing extant data for the well-studied system GU Boo, showing that our very different analysis methods match previous radius measurements to within 2--3\%. A similar result was found for GU Boo and yet another pipeline by \citet{Windmiller:2010rs}.  Given the wide variety of analysis methods used in literature studies of low-mass eclipsing systems, we suggest that interpretation of the stellar mass-radius relation should include a systematic term of 2--3\% unless all analyses were conducted with similar observations and analysis pipelines.

\section{Appendix A: Pipeline Validation via Reanalysis of GU Boo} \label{sec:guboo}

One of the first benchmark low-mass eclipsing binaries identified and rigorously characterized was GU Boo (\citealt{Lopez-Morales:2005jf}; hereafter LMR05), a pair of $\sim$0.6 $M_{\odot}$ (SpT = M0) stars in a short-period ($P = 0.48$ d) binary system. The radial velocities and multi-color photometry were published along with the system's discovery and characterization, making it a natural test case for comparing eclipsing binary analysis pipelines. In particular, LMR05 analyzed the system with different choices of software (the 2003 version of the Wilson-Devinney code; \citealt{Wilson:1971ig}), convergence algorithm (the standard Wilson-Devinney method of differential corrections), treatment of the atmospheres (via Kurucz models; e.g., \citealt{Kurucz:1979kl}), limb darkening (a square-root law), and treatment of spots (by modeling a large spot on each star).

To test the fidelity of our own pipeline, we have reanalyzed the RVs and optical photometry of GU Boo using the parts of our pipeline for which appropriate data were available. We specifically fit the individual component RVs and the $R_C$ and $I_C$ light curves, but did not attempt to apply any spectroscopic prior. We adopted the LMR05 assumption of component temperatures of $T_P = 3920$ K and $T_S= 3810$ K, and adopted quadratic limb darkening coefficients of $\gamma_{1,P,R} = 0.2516$, $\gamma_{2,P,R} = 0.3528$, $\gamma_{1,S,R} = 0.2562$, $\gamma_{2,S,R} = 0.4203$, $\gamma_{1,P,I} = 0.4558$, $\gamma_{2,P,I} = 0.3528$, $\gamma_{1,S,I} = 0.4763$, and $\gamma_{2,S,I} = 0.3829$. To rectify the secular influence of spots on the light curve, we fit and divided the phased light curves for near-eclipse observations (within $\Delta \phi < 0.15$) with quadratic polynomials; we otherwise did not include spot modeling. We fixed the orbital eccentricity to zero, and therefore do not fit $e$ or $\omega$. All other parameters were allowed to float, in analogy to our analysis of \ptfebb. We explored the posterior parameter distributions for this system with the MCMC pipeline described in Section~\ref{sec:analysis:mcmc}, computing 20 simultaneous chains for a total length of $1.1 \times 10^5$ steps per chain, omitting the first $10^4$ steps of each chain to allow for random disperal from the (common) initial starting point. We report the resulting system parameters in Table~\ref{tab:GUpartable}, and plot the resulting RV curves (Figure~\ref{fig:GUrvfit}), light curves (Figure~\ref{fig:GURfit} and Figure~\ref{fig:GUIfit}), and corner plot (Figure~\ref{fig:GUcorner}).

We find that our inferred component masses and the orbital parameters constrained by the RVs are indistinguishable from those of LMR05, as should be expected from fitting the same data with well-understood Newtonian dynamics. We do find modest differences in the parameters set by the light curves (stellar radii and system inclination), but our best-fit values agree with the values reported by LMR05 to within the mutual $1\sigma$ confidence interval. They found a system inclination of $i = 87.6\degr \pm 0.2\degr$, while our analysis finds a best-fit value of $i = 87.46\degr \pm 0.08\degr$ (0.75$\sigma$ lower). Similarly, they found stellar radii of $R_P = 0.623 \pm 0.016 R_{\odot}$ and $R_S = 0.620 \pm 0.020 R_{\odot}$, while we found radii of $R_P = 0.602 \pm 0.007 R_{\odot}$ and $R_S = .633 \pm 0.006 R_{\odot}$. We found that the sum of the radii was very well constrained ($R_{tot} = 1.235 \pm 0.003 R_{\odot}$), in good agreement with their sum ($R_{tot} = 1.243 \pm 0.026 R_{\odot}$), which confirms that the anti-correlated discrepancies are due to the well-known difficulty of measuring the ratio of radii in eclipsing binary systems. 

We see evidence of red (temporally correlated) noise in the light curves during eclipses, which we attribute to variations in surface brightness on the occulted star due to the presence of spots. In particular, the $R$ band light curve shows a clear asymmetry in the primary eclipse that might indicate a higher spot coverage fraction on the ingress side of the occulted primary star than on the egress side. This asymmetry qualitatively explains the location of the large spot that LMR05 added to the primary star in their system model.

In summary, these results suggest that even given the many methodological differences and absence of spot modeling, our pipeline yields radius measurements that are consistent with previous analysis pipelines to within $<$2--3\%.

\acknowledgements

~

We thank Phil Muirhead and Eunkyu Han for interesting discussions on low-mass eclipsing binaries and for comparing results on our respective pipelines, Trent Dupuy for helpful discussions on the fitting of orbits, and Lynne Hillenbrand for contributing to the Keck/HIRES data acquisition. We also thank the referee for providing a helpful and thoughtful critique of our paper that led to a number of valuable additions to the manuscript. A.W.M. was supported through Hubble Fellowship grant 51364 awarded by STScI, which is operated by AURA for NASA, under contract NAS 5-26555. GT acknowledges partial support for this work from NSF grant AST-1509375. M.A.A. acknowledges support provided by the NSF through grants AST- 1255419 and AST-1517367.

This paper includes data collected by the \Ktwo mission. Funding for the \Ktwo mission is provided by the NASA Science Mission directorate. The \Ktwo data presented in this paper were obtained from the Mikulski Archive for Space Telescopes. Funding for the Sloan Digital Sky Survey IV has been provided by the Alfred P. Sloan Foundation, the U.S. Department of Energy Office of Science, and the Participating Institutions. SDSS-IV acknowledges support and resources from the Center for High-Performance Computing at the University of Utah. The SDSS web site is www.sdss.org. SDSS-IV is managed by the Astrophysical Research Consortium for the Participating Institutions of the SDSS Collaboration.

This research also has made use of the Keck Observatory Archive (KOA), which is operated by the W. M. Keck Observatory and the NASA Exoplanet Science Institute (NExScI), under contract with the National Aeronautics and Space Administration. Some of the data presented herein were obtained at the W.M. Keck Observatory, which is operated as a scientific partnership among the California Institute of Technology, the University of California and the National Aeronautics and Space Administration. The Observatory was made possible by the generous financial support of the W.M. Keck Foundation. 

The authors wish to recognize and acknowledge the very significant cultural role and reverence that the summit of Mauna Kea has always had within the indigenous Hawaiian community.  We are most fortunate to have the opportunity to conduct observations from this mountain.

%\nocite{*}

\bibliographystyle{aasjournal.bst}
\bibliography{ms.bbl}
%\bibliography{../Papers/krausbib}

\clearpage

\floattable
\begin{deluxetable}{lcrl}
\tabletypesize{\footnotesize}
\tablewidth{0pt}
\tablecaption{PTF Photometry \label{tab:ptfphot}}
\tablehead{
\colhead{Epoch} & \colhead{Phase} & \colhead{$R$} & \colhead{$\sigma_R$}
\\
\colhead{(HJD-2450000)} & \colhead{} & \colhead{(mag)} & \colhead{(mag)}
}
\startdata
5229.7300 & 0.6192 & 16.987 &  0.060 \\
5229.7420 & 0.6212 & 16.910 &  0.038 \\
5229.7430 & 0.6214 & 16.951 &  0.042 \\
5229.7490 & 0.6224 & 17.035 &  0.044 \\
5229.7500 & 0.6226 & 16.923 &  0.048 \\
5229.7550 & 0.6234 & 16.969 &  0.050 \\
5229.7570 & 0.6237 & 17.015 &  0.060 \\
5229.7840 & 0.6282 & 16.919 &  0.052 \\
5229.7980 & 0.6305 & 16.992 &  0.054 \\
5229.8030 & 0.6314 & 17.057 &  0.043 \\
5229.8050 & 0.6317 & 16.941 &  0.049 \\
5229.8240 & 0.6349 & 17.029 &  0.045 \\
5229.8260 & 0.6352 & 16.996 &  0.044 \\
5239.7080 & 0.2779 & 17.070 &  0.025 \\
5239.7100 & 0.2782 & 17.100 &  0.024 \\
5239.7210 & 0.2800 & 17.052 &  0.024 \\
5239.7230 & 0.2804 & 17.083 &  0.027 \\
5239.7330 & 0.2820 & 16.985 &  0.037 \\
5239.7710 & 0.2884 & 17.075 &  0.049 \\
5239.7730 & 0.2887 & 17.087 &  0.062 \\
5239.8040 & 0.2938 & 17.040 &  0.024 \\

\enddata
\tablecomments{The full table is available as a machine-readable table, ptfphotFULL1111.txt.}
\end{deluxetable}

\floattable
\begin{deluxetable}{lccc}
\tabletypesize{\footnotesize}
\tablewidth{0pt}
\tablecaption{\Ktwo Photometry \label{tab:k2phot}}
\tablehead{
\colhead{Epoch} & \colhead{Phase} & \colhead{$F$} & \colhead{$\sigma_F$}\\
\colhead{(HJD-2450000)} & \colhead{} & \colhead{(normalized)} & \colhead{(normalized)}
}
\startdata
7139.6011 & 0.0981 &  0.995 &  0.008 \\
7139.6215 & 0.1015 &  0.999 &  0.007 \\
7139.6419 & 0.1049 &  0.995 &  0.007 \\
7139.6624 & 0.1083 &  0.994 &  0.007 \\
7139.6828 & 0.1117 &  1.002 &  0.007 \\
7139.7032 & 0.1151 &  0.998 &  0.007 \\
7139.7236 & 0.1185 &  1.001 &  0.007 \\
7139.7441 & 0.1219 &  1.015 &  0.007 \\
7139.7645 & 0.1253 &  1.018 &  0.007 \\
7139.7849 & 0.1287 &  1.004 &  0.007 \\
7139.8054 & 0.1321 &  0.996 &  0.007 \\
7139.8258 & 0.1355 &  0.977 &  0.007 \\
7139.8462 & 0.1389 &  0.860 &  0.007 \\
7139.8667 & 0.1423 &  0.799 &  0.007 \\
7139.8871 & 0.1456 &  0.882 &  0.007 \\
7139.9075 & 0.1490 &  0.993 &  0.007 \\
7139.9280 & 0.1524 &  1.000 &  0.007 \\
7139.9484 & 0.1558 &  0.998 &  0.007 \\
7139.9688 & 0.1592 &  0.995 &  0.007 \\
7139.9893 & 0.1626 &  1.000 &  0.007 \\
7140.0097 & 0.1660 &  0.998 &  0.007 \\

\enddata
\tablecomments{The full table is available as a machine-readable table, k2photFULL1111.txt.}
\end{deluxetable}

\floattable
\begin{deluxetable}{lrl}
\tabletypesize{\footnotesize}
\tablewidth{0pt}
\tablecaption{System Photometry \label{tab:catphot}}
\tablehead{
\colhead{Filter} & \colhead{$m$ (mag)} & \colhead{Reference}
}
\startdata
$u$ & 21.021 $\pm$ 0.080 & SDSS-DR9 \citep{Ahn:2012ys}\\
$g$ & 18.771 $\pm$ 0.008 & SDSS-DR9 \citep{Ahn:2012ys}\\
$r$ & 17.302 $\pm$ 0.006 & SDSS-DR9 \citep{Ahn:2012ys}\\
$i$ & 15.807 $\pm$ 0.004 & SDSS-DR9 \citep{Ahn:2012ys}\\
$z$ & 14.999 $\pm$ 0.005 & SDSS-DR9 \citep{Ahn:2012ys}\\
$J$ & 13.529 $\pm$ 0.026 & 2MASS \citep{Cutri:2003jh} \\
$H$ & 12.911 $\pm$ 0.024 & 2MASS \citep{Cutri:2003jh} \\
$K_s$ & 12.651 $\pm$ 0.022 & 2MASS \citep{Cutri:2003jh} \\
$W1$ & 12.497 $\pm$ 0.024 & ALLWISE \citep{Cutri:2013lq}\\
$W2$ & 12.330 $\pm$ 0.024 & ALLWISE \citep{Cutri:2013lq}\\
\enddata
\end{deluxetable}

\floattable
\begin{deluxetable*}{lccrrrrrr}
\tabletypesize{\footnotesize}
\tablewidth{0pt}
\tablecaption{Keck-I/HIRES RVs \label{tab:hiresRV}} 
\tablehead{
\colhead{Target/} & \colhead{Epoch} & \colhead{Wavelength} &  \colhead{$t_{int}$)} &  \colhead{$v_{p}$} & \colhead{$\sigma_{v_p}$\tablenotemark{b}} & \colhead{$v_{s}$} & \colhead{$\sigma_{v_s}$\tablenotemark{b}} & \colhead{$F_s/F_p$}
\\
\colhead{Epoch\tablenotemark{a}} & \colhead{(HJD-2450000)} & \colhead{Range (\AA)} & \colhead{(sec)} & \colhead{(km/s)} & \colhead{(km/s)} & \colhead{(km/s)} & \colhead{(km/s)} 
}
\startdata
      20101213.36161 & 5543.92230 & 4450--8910 &  600 &    1.53$\pm$ 0.08 &    7.84$\pm$ 0.10 &   96.38$\pm$ 0.22 &    7.45$\pm$ 0.27 &  0.259$\pm$ 0.013 \\ %   Flux-071
      20101213.42871 & 5543.99997 & 4450--8910 &  600 &    1.00$\pm$ 0.07 &    7.71$\pm$ 0.07 &   97.61$\pm$ 0.23 &    7.78$\pm$ 0.30 &  0.271$\pm$ 0.013 \\ %   Flux-086
      20101213.46220 & 5544.03873 & 4450--8910 &  600 &    0.42$\pm$ 0.09 &    7.79$\pm$ 0.07 &   97.74$\pm$ 0.23 &    7.48$\pm$ 0.24 &  0.279$\pm$ 0.011 \\ %   Flux-090
      20101213.49879 & 5544.08108 & 4450--8910 &  450 &    0.29$\pm$ 0.11 &    7.77$\pm$ 0.10 &   98.72$\pm$ 0.26 &    7.50$\pm$ 0.26 &  0.274$\pm$ 0.017 \\ %   Flux-098
      20120104.40672 & 5930.97583 & 4450--8910 &  600 &   44.04$\pm$ 0.10 &    7.77$\pm$ 0.09 &   16.96$\pm$ 0.18 &    7.60$\pm$ 0.16 &  0.283$\pm$ 0.013 \\ %  Flux-086b
      20120104.46819 & 5931.04698 & 4450--8910 &  900 &   46.56$\pm$ 0.10 &    7.84$\pm$ 0.08 &   12.31$\pm$ 0.14 &    7.84$\pm$ 0.22 &  0.266$\pm$ 0.012 \\ %  Flux-098b
      20120104.55549 & 5931.14802 & 4450--8910 &  600 &   49.54$\pm$ 0.11 &    7.73$\pm$ 0.08 &    6.10$\pm$ 0.26 &    7.71$\pm$ 0.48 &  0.268$\pm$ 0.012 \\ %   Flux-114
      20120106.39276 & 5932.95976 & 4450--8910 &  600 &   59.20$\pm$ 0.11 &    7.71$\pm$ 0.10 &  -11.52$\pm$ 0.15 &    7.49$\pm$ 0.22 &  0.279$\pm$ 0.011 \\ %   Flux-092
      20120106.47033 & 5933.04954 & 4450--8910 & 1200 &   56.81$\pm$ 0.10 &    7.83$\pm$ 0.06 &   -6.78$\pm$ 0.19 &    7.39$\pm$ 0.12 &  0.261$\pm$ 0.012 \\ %   Flux-109
      20120106.55325 & 5933.14552 & 4450--8910 &  600 &   53.91$\pm$ 0.12 &    7.84$\pm$ 0.07 &   -2.04$\pm$ 0.30 &    7.44$\pm$ 0.25 &  0.279$\pm$ 0.014 \\ %   Flux-119
      20100526.23134 & 5342.76526 & 3360--8100 &  900 &   68.01$\pm$ 0.19 &    8.02$\pm$ 0.14 &  -31.24$\pm$ 0.49 &    8.13$\pm$ 0.55 &  0.263$\pm$ 0.017 \\ %     j91.84
      20100603.22375 & 5350.75579 & 3360--8100 &  900 &   14.87$\pm$ 0.11 &    7.89$\pm$ 0.13 &   69.87$\pm$ 0.34 &    7.35$\pm$ 0.48 &  0.229$\pm$ 0.023 \\ %     j92.78
      20101121.44093 & 5522.01223 & 3360--8100 &  450 &   45.15$\pm$ 1.52 &    8.92$\pm$ 0.87 &    8.88$\pm$ 0.43 &    7.72$\pm$ 0.20 &  0.291$\pm$ 0.025 \\ %   j106.278
      20101122.52457 & 5523.10914 & 3360--8100 &  450 &   67.70$\pm$ 0.22 &    7.92$\pm$ 0.20 &  -29.52$\pm$ 0.85 &    8.68$\pm$ 0.88 &  0.312$\pm$ 0.027 \\ %   j106.606
      20101212.36604 & 5542.92735 & 3360--8100 &  600 &   24.98$\pm$ 0.12 &    7.62$\pm$ 0.16 &   48.88$\pm$ 0.53 &    8.63$\pm$ 0.49 &  0.292$\pm$ 0.022 \\ %   j109.126
      20101212.47686 & 5543.05563 & 3360--8100 &  600 &   20.94$\pm$ 0.12 &    7.69$\pm$ 0.09 &   57.21$\pm$ 0.36 &    7.57$\pm$ 0.31 &  0.273$\pm$ 0.018 \\ %   j109.167
      20101214.52722 & 5545.11407 & 3360--8100 &  600 &   13.12$\pm$ 0.14 &    7.71$\pm$ 0.14 &   70.18$\pm$ 0.69 &    8.98$\pm$ 0.71 &  0.335$\pm$ 0.030 \\ %   j110.157
      20101215.43555 & 5546.00803 & 3360--8100 &  600 &   43.74$\pm$ 0.16 &    7.95$\pm$ 0.17 &   12.51$\pm$ 0.40 &    7.82$\pm$ 0.59 &  0.324$\pm$ 0.032 \\ %   j110.544
      20110123.41492 & 5584.98588 & 3360--8100 &  600 &   26.75$\pm$ 0.16 &    7.99$\pm$ 0.23 &   45.95$\pm$ 1.12 &    6.73$\pm$ 0.86 &  0.219$\pm$ 0.041 \\ %   j113.167
      20120105.39850 & 5931.96637 & 3360--8100 &  600 &   66.37$\pm$ 0.13 &    7.82$\pm$ 0.11 &  -28.96$\pm$ 0.42 &    7.41$\pm$ 0.28 &  0.276$\pm$ 0.016 \\ %   j141.134

\hline
Gl 83.1 & 5741.13595 & 4320-8750 & 120 & .. & .. & .. & .. & .. \\
Gl 83.1 & 5930.69346 & 4320-8750 & 120 & .. & .. & .. & .. & ..  \\
Gl 447 & 5933.16968 & 4320-8750 & 120 & .. & .. & .. & .. & ..  \\
HZ 44 & 5931.18067 & 4320-8750 & 120 & .. & .. & .. & .. & .. \\
\enddata
\tablecomments{In each observation, the component velocities are subject to a shared systematic uncertainty of $\pm 300$ m/s from the uncertainty in the absolute RV scale. Furthermore, the velocities at all epochs are subject to a shared systematic uncertainty of $\pm 170$ m/s because they are all measured with respect to the same three calibrator stars, each of which has a systematic uncertainty of $\pm$300 m/s.}
\tablenotetext{a}{The first column lists either the UT date and time stamp from the Keck Observatory Archive (for observations of \ptfebb) or the target name (for standard stars).}
\tablenotetext{b}{We report $\sigma_{v_p}$ and $\sigma_{v_s}$ as the standard deviation of the Gaussian fits to the two stars' broadening functions, which is a measure of both the instrumental broadening and the rotational broadening. We discuss the conversion to $v \sin(i)$ in Section~\ref{sec:obs:hires}.}
\end{deluxetable*}

\floattable
\begin{deluxetable}{cccc}
\tabletypesize{\footnotesize}
\tablewidth{0pt}
\tablecaption{H$\alpha$ Equivalent Widths Near Quadrature \label{tab:ewha}}
\tablehead{
\colhead{Epoch} & \colhead{Phase} & \colhead{$EW[H\alpha]_{P}$} & \colhead{$EW[H\alpha]_{S}$}
\\
\colhead{(HJD-2450000)} & \colhead{} & \colhead{(\AA)} & \colhead{(\AA)}
}
\startdata
55342.765 & 0.414 & 3.09 & 0.50 \\
55523.109 & 0.392 & 2.70 & 0.52 \\
55543.922 & 0.852 & 3.46 & 0.58 \\
55544.000 & 0.865 & 3.63 & 0.46 \\
55544.039 & 0.871 & 3.47 & 0.56 \\
55544.081 & 0.878 & 3.48 & 0.45 \\
55931.966 & 0.357 & 3.27 & 0.45 \\

\enddata
\end{deluxetable}

 \begin{figure*}
 \epsscale{0.95}
\includegraphics[scale=0.5]{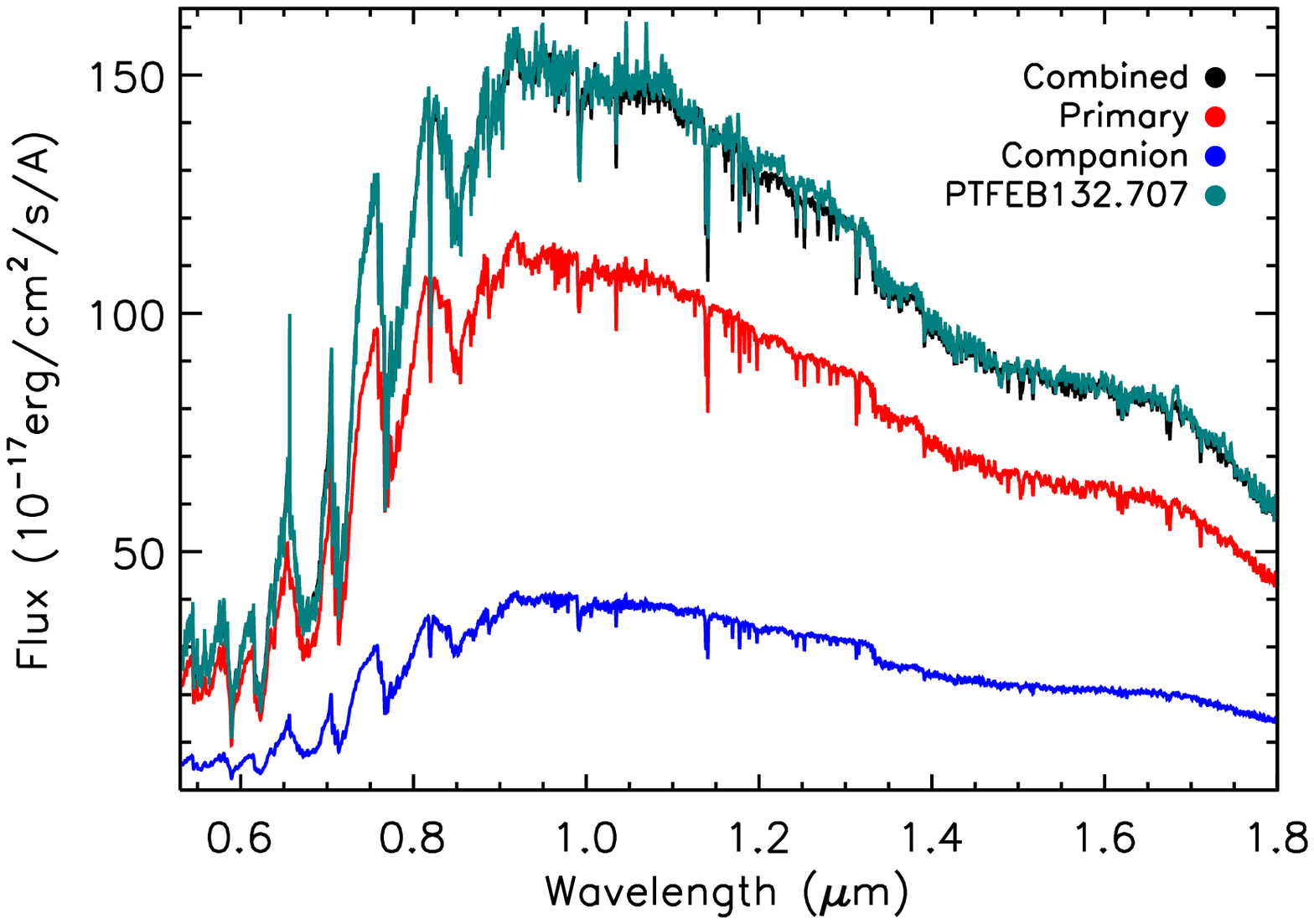}
\includegraphics[scale=0.5]{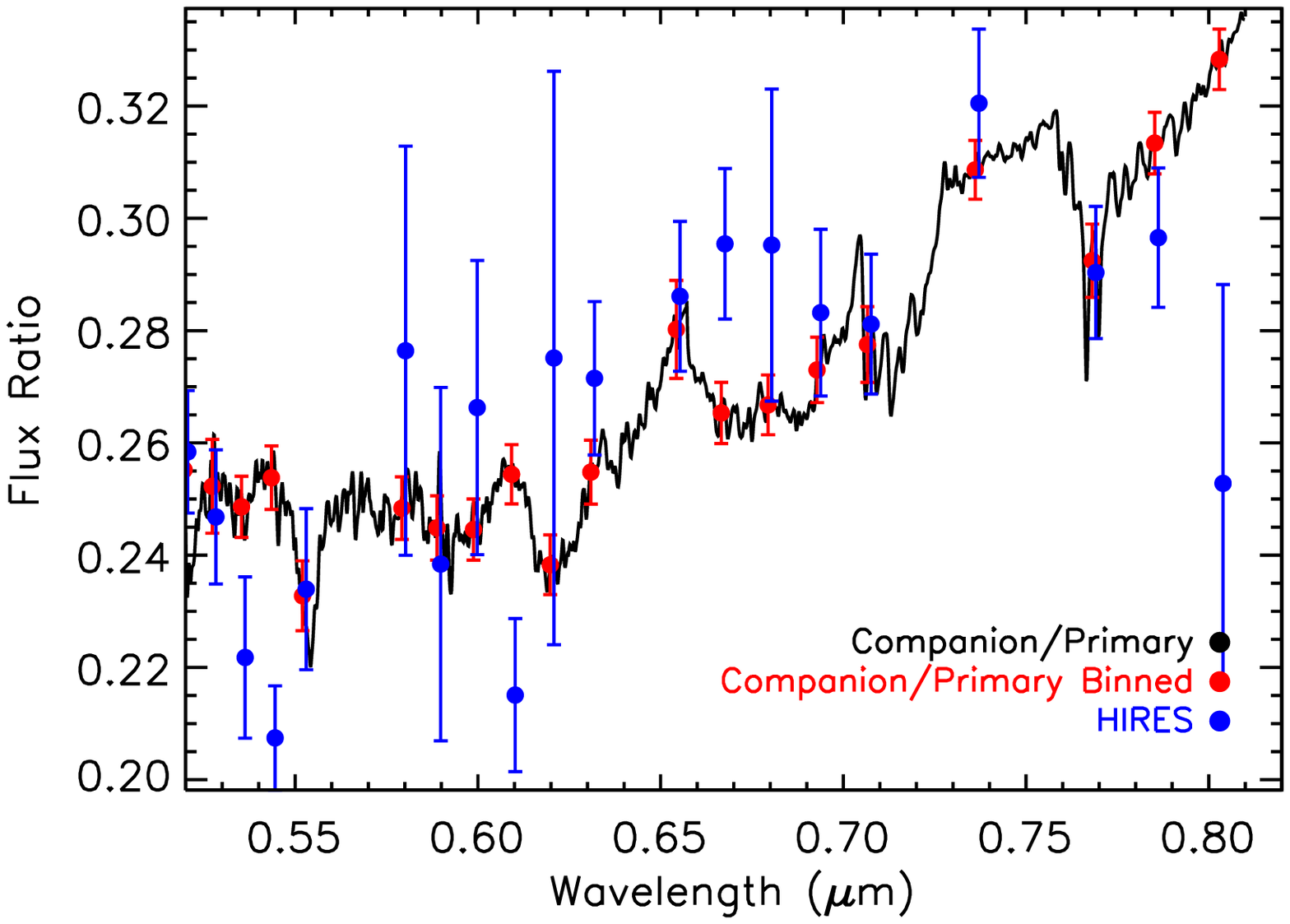}
 \caption{Results of our SED fitting procedure. Left: Unresolved SNIFS+SpeX composite spectrum for \ptfeb (teal), the best-fitting template spectra (red and blue), and their sum (black). Right: Flux ratios measured in each order of the Keck/HIRES spectra (blue points), the ratio of the two best-fitting component spectra (black line), and the corresponding binned flux ratio values for direct comparison (red points).
 \label{fig:sedfit}} 
  \end{figure*}

\begin{figure*}
 \epsscale{0.95}
\includegraphics[scale=0.43]{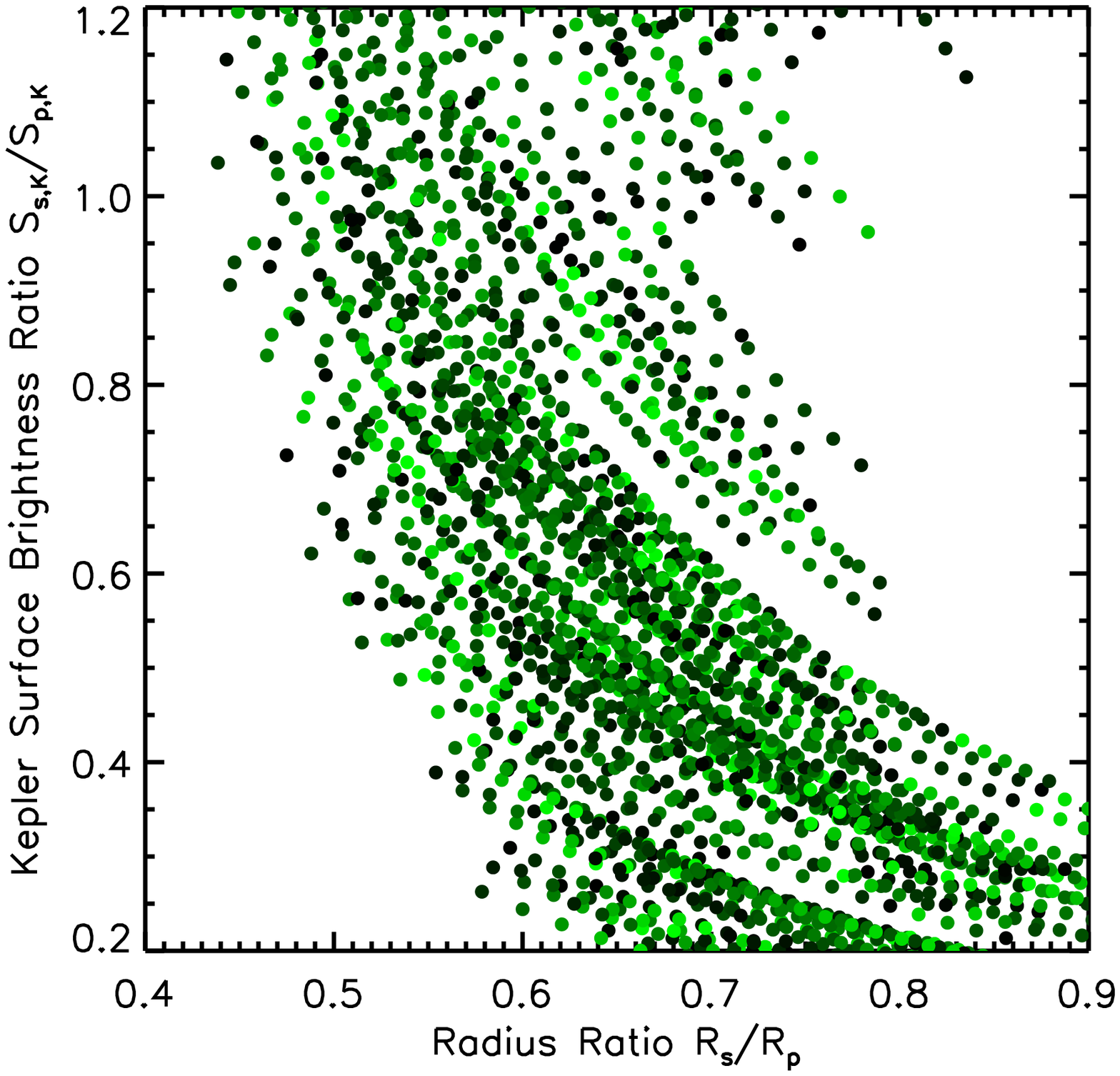}
\includegraphics[scale=0.43]{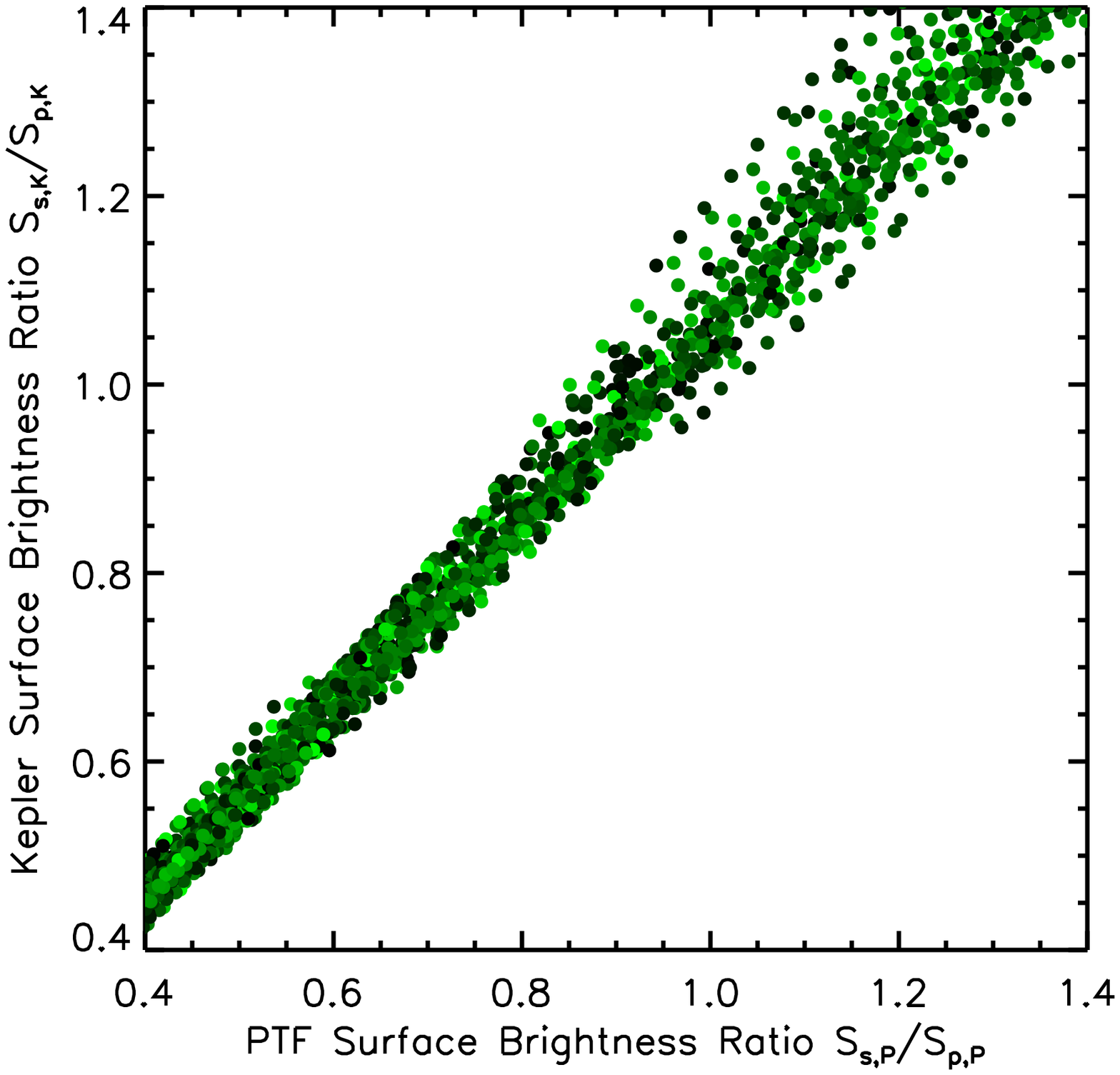}
 \caption{Posterior distribution from our SED fitting procedure. We show the results for all pairs of template spectra that yielded $\chi_{\nu}^2 < 4$, plotting the resulting normalization parameters in the 3D posterior as projected into two planes (left: $\frac{R_s}{R_p}$ vs $\frac{S_{s,Kep}}{S_{p,Kep}}$; right: $\frac{S_{s,Kep}}{S_{p,Kep}}$ vs $\frac{S_{s,PTF}}{S_{p,PTF}}$). The points are shaded green, with hue corresponding to the $\chi_{\nu}^2$ value for that pair of templates, ranging from 1 (bright green) to 4 (black). The discrete tracks denote points resulting from the same pair of spectra that were combined with a different normalization; typically a range of normalizations resulted in acceptable $\chi^2$ values. We take the density of these points in parameter space as a proxy for the true posterior, since several confounding variables (such as metallicity, age, and spot coverage) prevent the SED fit results from defining a simple $\chi^2$ hypersurface. \label{fig:RFFpost}} 
\end{figure*}

\clearpage

\floattable
\begin{deluxetable}{lr}
\tabletypesize{\footnotesize}
\tablewidth{0pt}
\tablecaption{System Parameters for \ptfeb \label{tab:partable}}
\tablehead{}
\startdata
\hline \multicolumn{2}{l}{\it Orbital Parameters}\\ \hline
                          $T_0$ (HJD) &    2457145.0    $\pm$            0.4 \\
                          $T_P$ (HJD) &    2457148.9041 $\pm$         0.0001 \\
                           $P$ (days) &        6.015742 $\pm$       0.000002 \\
                             $a$ (AU) &         0.05475 $\pm$        0.00006 \\
                                  $e$ &         0.0017  $\pm$         0.0006 \\
                            $i$ (deg) &          88.87  $\pm$           0.05 \\
                       $\omega$ (deg) &          38     $\pm$             27 \\
                      $\gamma$ (km/s) &           34.00 $\pm$           0.15 \\
\hline \multicolumn{2}{l}{Stellar Bulk Parameters}\\ \hline
              $M_p+M_s$ ($M_{\odot}$) &          0.6050 $\pm$         0.0020 \\
                        $q = M_s/M_p$ &           0.531 $\pm$          0.005 \\
                  $M_p$ ($M_{\odot}$) &          0.3953 $\pm$         0.0020 \\
                  $M_s$ ($M_{\odot}$) &          0.2098 $\pm$         0.0014 \\
              $R_p+R_s$ ($R_{\odot}$) &          0.635  $\pm$          0.005 \\
                            $R_s/R_p$ &           0.75  $\pm$           0.05 \\
                  $R_p$ ($R_{\odot}$) &          0.363  $\pm$          0.008 \\
                  $R_s$ ($R_{\odot}$) &          0.272  $\pm$          0.012 \\
\hline \multicolumn{2}{l}{Stellar Atmospheric Parameters}\\ \hline
                    $S_{s,K}/S_{p,K}$ &           0.699 $\pm$          0.006 \\
                    $S_{s,P}/S_{p,P}$ &           0.66  $\pm$           0.04 \\

\hline \multicolumn{2}{l}{Unresolved Stellar Parameters}\\ \hline
                   $F_{bol}$ (erg/s/cm$^2$) &    $(1.75 \pm 0.06) \times 10^{-11}$ \\
                   $L_{bol}$ ($L_{\odot}$) &    $0.0180 \pm 0.0010$ \\
\hline \multicolumn{2}{l}{Primary Star Parameters}\\ \hline
                                  SpT &            M3.5 $\pm$ 0.2 $\pm$ 0.3 \\
                    $T_{\rm eff}$ (K) &            3260 $\pm$ 30 $\pm$ 60\\
                   $F_{bol}$ (erg/s/cm$^2$) &    $(1.32 \pm 0.05) \times 10^{-11}$ ($\pm$2\%)\\
                   $L_{bol}$ ($L_{\odot}$) &    $0.0137 \pm 0.0010$ \\
\hline \multicolumn{2}{l}{Secondary Star Parameters}\\ \hline
                                  SpT &            M4.3 $\pm$ 0.7 $\pm$0.3 \\
                    $T_{\rm eff}$ (K) &            3120 $\pm$ 50 $\pm$60 \\
                   $F_{bol}$ (erg/s/cm$^2$) &    $(0.49 \pm 0.06) \times 10^{-11}$ ($\pm$2\%)\\
                   $L_{bol}$ ($L_{\odot}$) &    $0.0050 \pm 0.0015$ \\
\enddata
\tablecomments{In all cases we report the median of the marginalized distribution. The values of $T_0$ and $\omega$ are individually poorly constrained, but are subject to a tight joint constraint that is captured by the time of primary eclipse $T_{P}$. To predict observations from the orbital elements, $\omega$ and $T_{P}$ should be used to compute an appropriate value of $T_0$ with sufficient precision. If $\omega$, $T_P$, $e$, and $P$ are fixed to the values listed in this table, $T_0 = 2457145.0267$. There is a small (4\%) difference between the system $L_{bol}$ and the sum of the component $L_{bol}$ because they are determined from different analysis methods.}
%                          $T_0$ (HJD) &    2457145.0267 $\pm$         0.4446 \\
%                       $\omega$ (deg) &          38.114 $\pm$         26.611 \\
\end{deluxetable}

 \begin{figure*}
 \epsscale{1.12}
 \includegraphics[scale=0.8,trim=0.0cm 0.0cm 0.0cm 0.0cm]{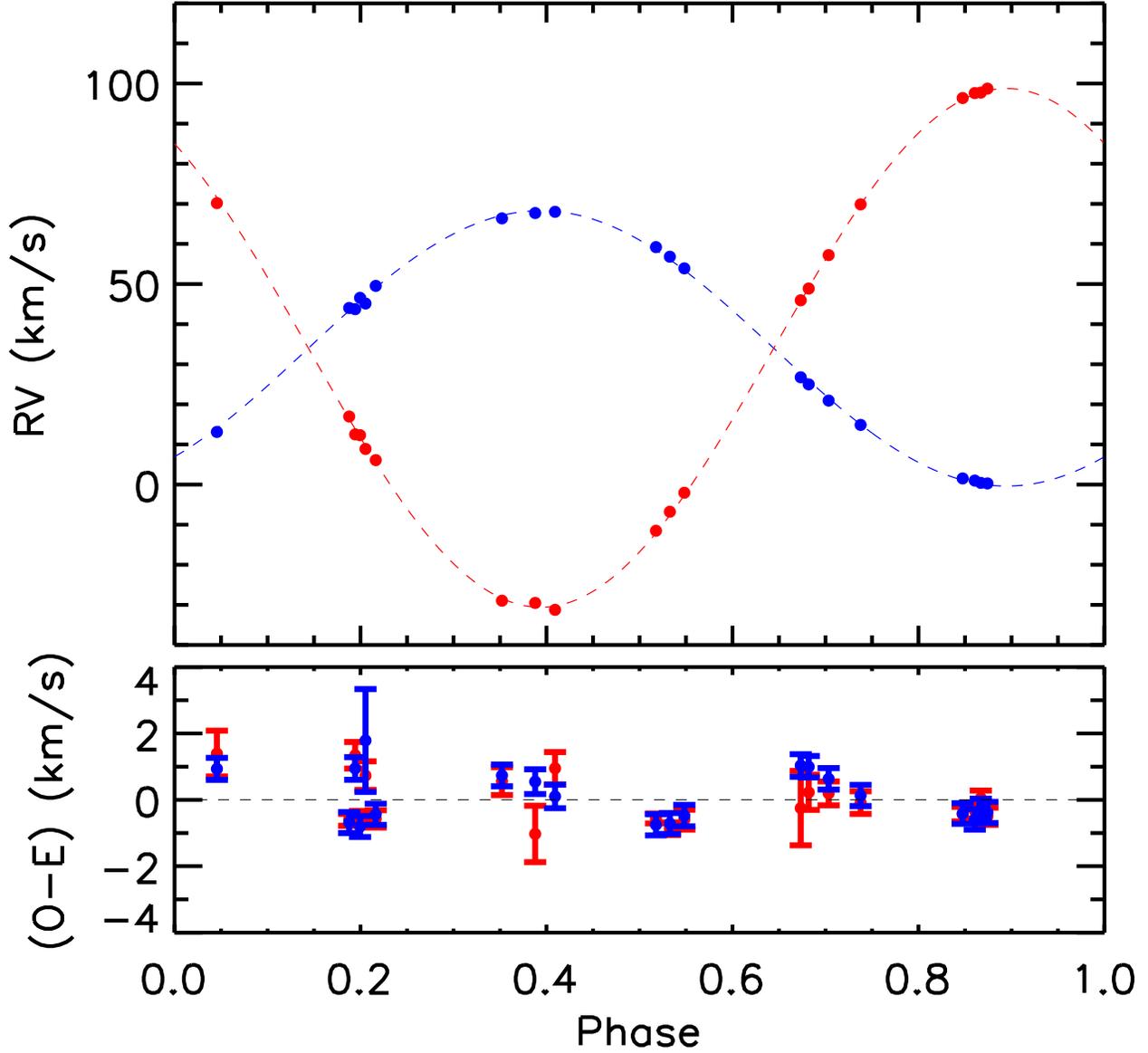}
 \caption{Radial velocities $v_p$ (blue) and $v_s$ (red) for the primary and secondary stars of \ptfebb, as measured from the Keck/HIRES epochs listed in Table~\ref{tab:hiresRV}. We also show the best-fit model as determined from our fitting procedure (Section~\ref{sec:results:sysprops}). Underneath, we show the (O-E) residuals with respect to the best-fit model. \label{fig:rvfit}} 
  \end{figure*}

   \begin{figure*}
 \epsscale{1.12}
 \plotone{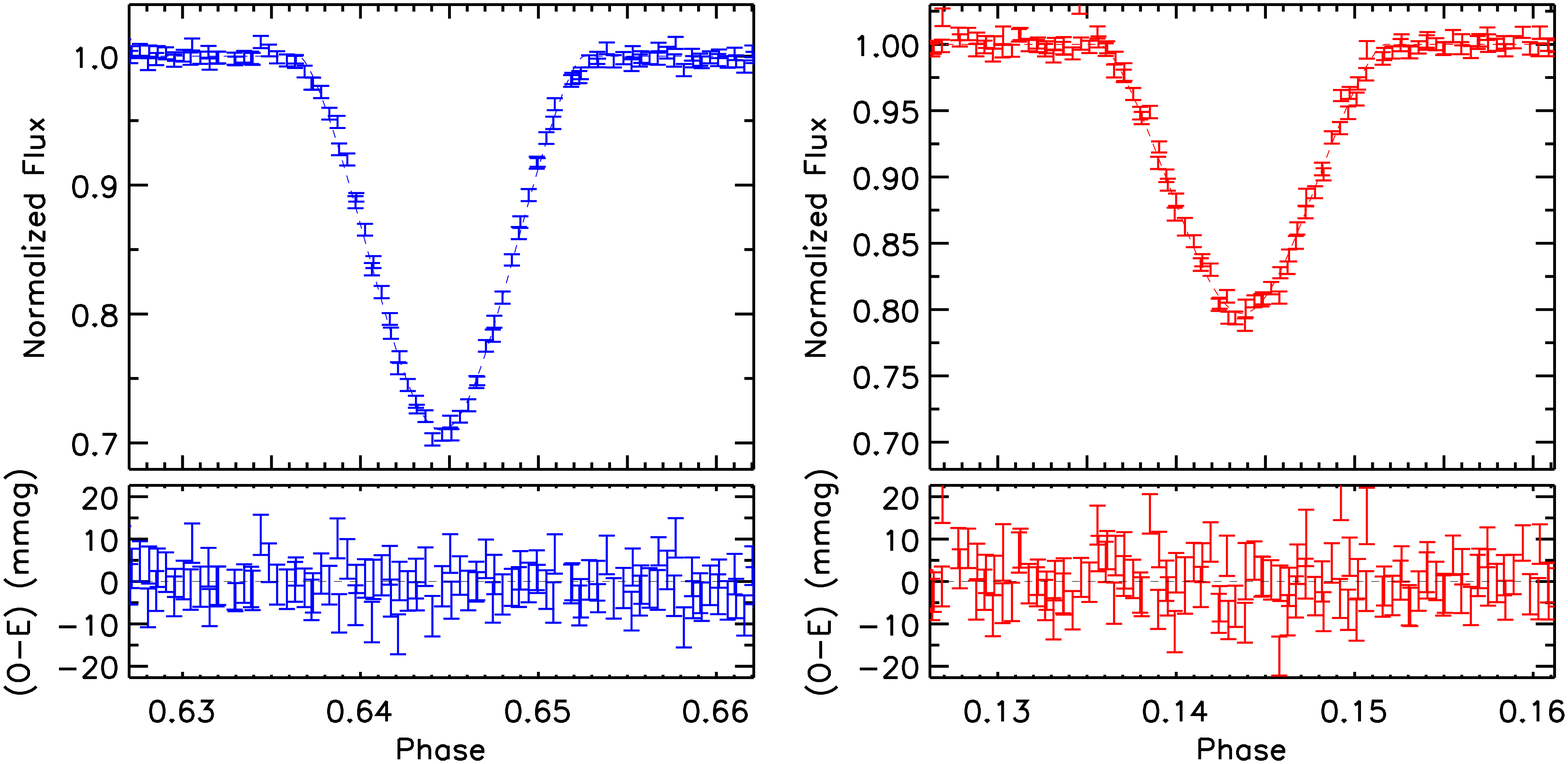}
 \caption{\Ktwo photometry for the primary eclipse (left) and secondary eclipse (right) of \ptfebb, along with the best-fitting models (dashed lines) and the (O-E) residuals (bottom panels). \label{fig:lcfit}} 
  \end{figure*}

     \begin{figure*}
 \epsscale{1.12}
 \plotone{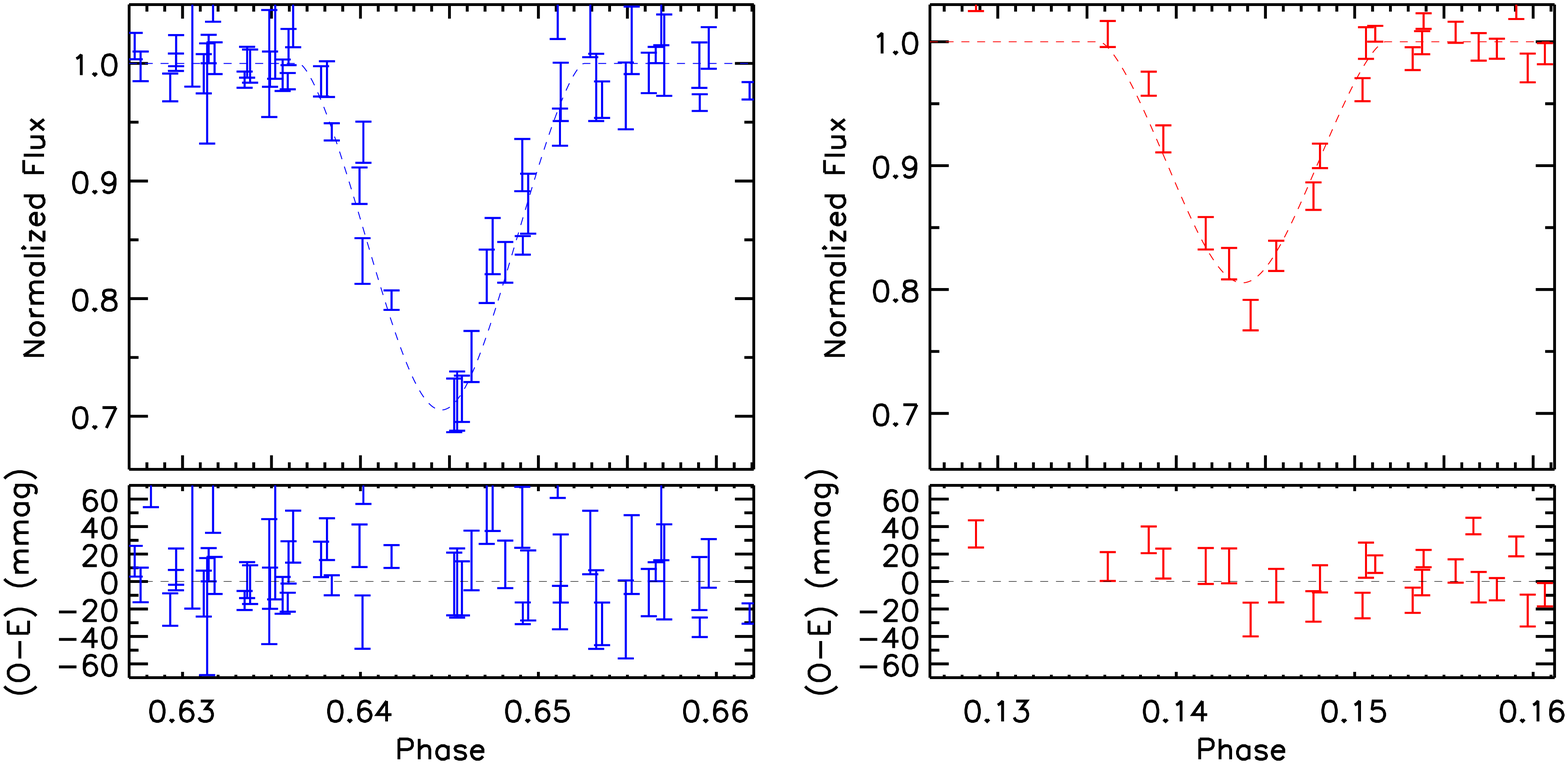}
 \caption{PTF photometry for the primary eclipse (left) and secondary eclipse (right) of \ptfebb, along with the best-fitting models (dashed lines) and the (O-E) residuals (bottom panels). \label{fig:ptffit}} 
  \end{figure*}

   \begin{figure*}
 \epsscale{1.12}
 \includegraphics[scale=0.44,trim=0.0cm 0.0cm 0.0cm 0.0cm]{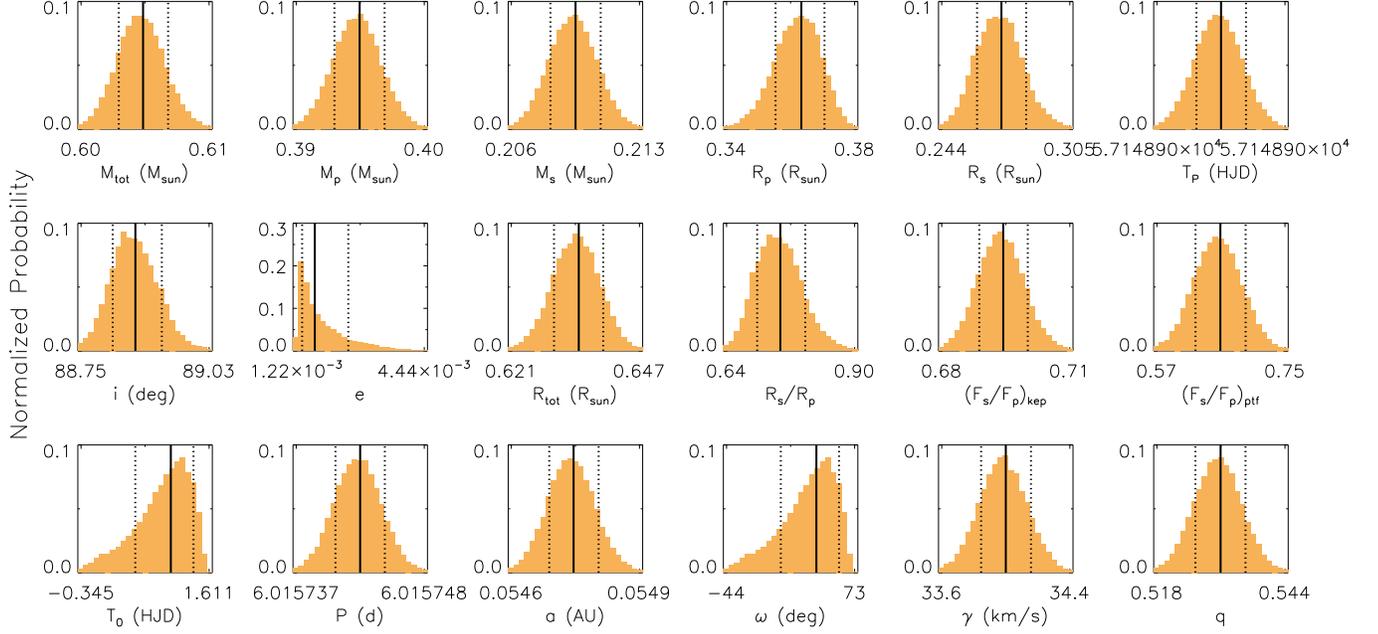}
 \caption{One-dimensional marginalized posterior distributions for the 12 fit parameters and six derived parameters in our MCMC. For each parameter, a solid black vertical line shows the median, while vertical dotted lines show the central 68\% credible interval. \label{fig:posteriors1d}} 
  \end{figure*}

   \begin{figure*}
 \epsscale{1.12}
 \includegraphics[scale=0.44,trim=0.0cm 0.0cm 0.0cm 0.0cm]{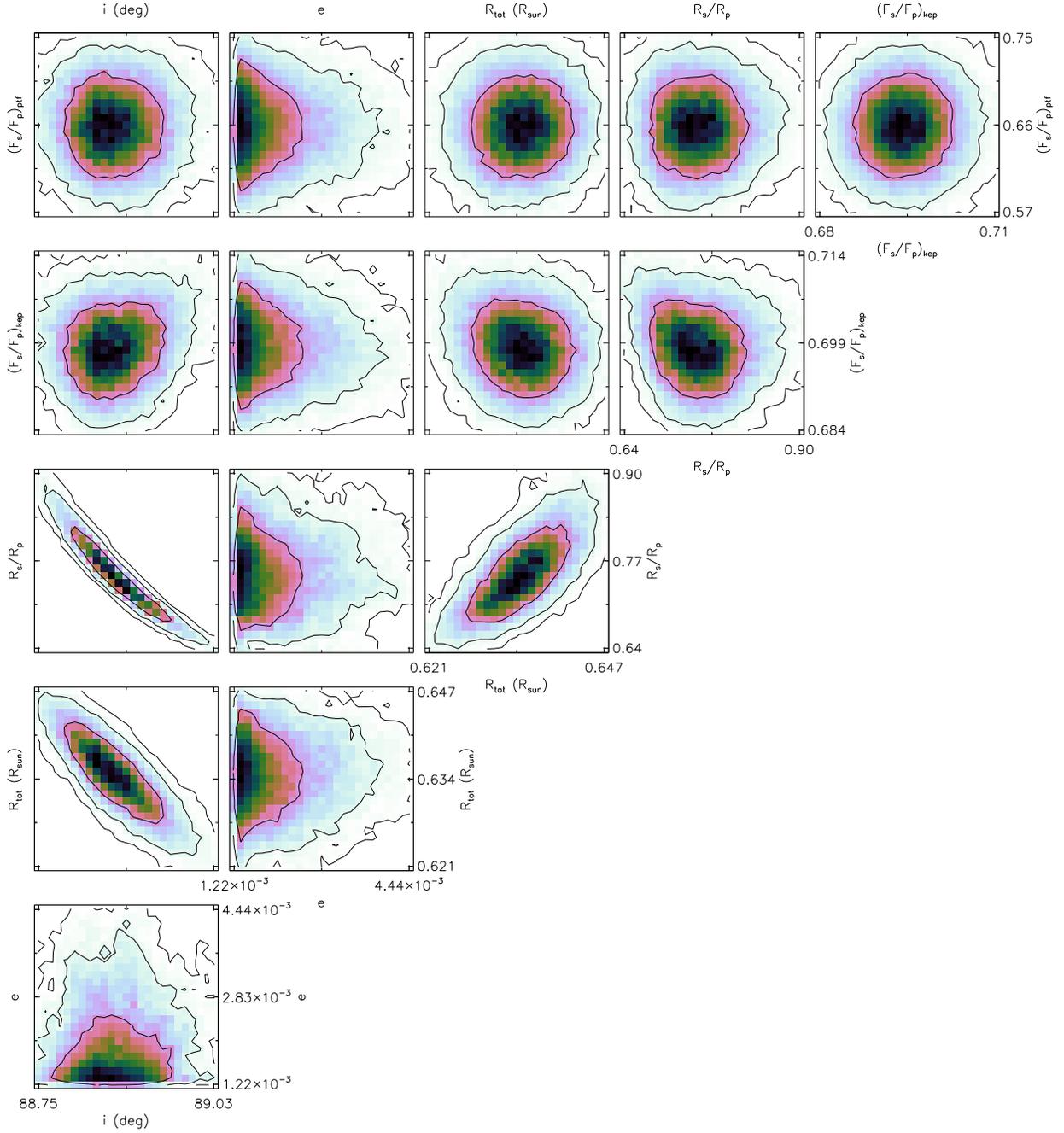}
 \caption{Two-dimensional marginalized posterior distributions for the 6 fit parameters in our MCMC that are most likely to show astrophysically important covariances. There is a strong covariance between the radius ratio, radius sum, and system inclination (column 1, rows 3--4 and column 3, row 3), and weaker covariance with the \Kepler bandpass surface brightness ratio (column 4, row 2). \label{fig:corner}} 
  \end{figure*}

\floattable
\begin{deluxetable}{lcccccccc}
\tabletypesize{\tiny}
\tablewidth{0pt}
\tablecaption{Variations in System Properties Derived from Data Subsets \label{tab:comptable}}
\tablehead{\colhead{Data} & \colhead{$e$} & \colhead{$i$} & \colhead{$R_p+R_s$} & \colhead{$\frac{R_s}{R_p}$} & \colhead{$R_p$} & \colhead{$R_s$} & \colhead{$\frac{S_{s,K}}{S_{p,K}}$} & \colhead{$\frac{S_{s,P}}{S_{p,P}}$}
\\
\colhead{} & \colhead{} & \colhead{(deg)} & \colhead{($R_{\odot}$)} & \colhead{} & \colhead{($R_{\odot}$)} & \colhead{($R_{\odot}$)}
}
\startdata
                           Full Fit &         0.00168 $\pm$        0.00058 &          88.871 $\pm$          0.054 &          0.6348 $\pm$         0.0051 &           0.751 $\pm$          0.048 &          0.3626 $\pm$         0.0080 &          0.2724 $\pm$         0.0115 &           0.699 $\pm$          0.006 &           0.659 $\pm$          0.036 \\
                      No Spec Prior &         0.00189 $\pm$        0.00070 &          88.820 $\pm$          0.084 &          0.6380 $\pm$         0.0064 &           0.806 $\pm$          0.104 &          0.3533 $\pm$         0.0168 &          0.2849 $\pm$         0.0228 &           0.700 $\pm$          0.007 &           0.676 $\pm$          0.043 \\
                             No PTF &         0.00158 $\pm$        0.00049 &          88.874 $\pm$          0.067 &          0.6340 $\pm$         0.0057 &           0.750 $\pm$          0.056 &          0.3626 $\pm$         0.0091 &          0.2718 $\pm$         0.0138 &           0.699 $\pm$          0.006 &           0.622 $\pm$          0.083 \\
                      No K2 Primary &         0.00177 $\pm$        0.00064 &          88.863 $\pm$          0.098 &          0.6287 $\pm$         0.0109 &           0.711 $\pm$          0.054 &          0.3676 $\pm$         0.0078 &          0.2613 $\pm$         0.0153 &           0.747 $\pm$          0.030 &           0.676 $\pm$          0.040 \\
                    No K2 Secondary &         0.00178 $\pm$        0.00063 &          88.891 $\pm$          0.087 &          0.6365 $\pm$         0.0065 &           0.688 $\pm$          0.056 &          0.3774 $\pm$         0.0101 &          0.2601 $\pm$         0.0147 &           0.623 $\pm$          0.061 &           0.652 $\pm$          0.036 \\
               No Spec Prior or PTF &         0.00154 $\pm$        0.00038 &          88.761 $\pm$          0.078 &          0.6398 $\pm$         0.0057 &           0.899 $\pm$          0.186 &          0.3377 $\pm$         0.0295 &          0.3038 $\pm$         0.0337 &           0.704 $\pm$          0.010 &           ... \\
        No Spec Prior or K2 Primary &         0.00166 $\pm$        0.00068 &          88.708 $\pm$          0.126 &          0.6434 $\pm$         0.0133 &           0.909 $\pm$          0.329 &          0.3386 $\pm$         0.0460 &          0.3076 $\pm$         0.0579 &           0.773 $\pm$          0.063 &           0.690 $\pm$          0.046 \\
      No Spec Prior or K2 Secondary &         0.00178 $\pm$        0.00066 &          88.845 $\pm$          0.122 &          0.6392 $\pm$         0.0080 &           0.696 $\pm$          0.093 &          0.3775 $\pm$         0.0167 &          0.2625 $\pm$         0.0235 &           0.551 $\pm$          0.088 &           0.681 $\pm$          0.044 \\
               No PTF or K2 Primary &         0.00108 $\pm$        0.00094 &          88.992 $\pm$          0.274 &          0.6152 $\pm$         0.0165 &           0.671 $\pm$          0.070 &          0.3667 $\pm$         0.0175 &          0.2464 $\pm$         0.0176 &           0.720 $\pm$          0.182 &           0.658 $\pm$          0.207 \\
             No PTF or K2 Secondary &         0.00088 $\pm$        0.00081 &          88.850 $\pm$          0.138 &          0.6385 $\pm$         0.0083 &           0.680 $\pm$          0.062 &          0.3790 $\pm$         0.0116 &          0.2582 $\pm$         0.0165 &           0.523 $\pm$          0.200 &           0.458 $\pm$          0.210 \\
      No K2 Primary or K2 Secondary &         0.00179 $\pm$        0.00086 &          88.843 $\pm$          0.152 &          0.6472 $\pm$         0.0232 &           0.673 $\pm$          0.079 &          0.3867 $\pm$         0.0175 &          0.2606 $\pm$         0.0236 &           0.744 $\pm$          0.104 &           0.679 $\pm$          0.045 \\
                           Only PTF &         0.00172 $\pm$        0.00099 &          88.638 $\pm$          0.139 &          0.6619 $\pm$         0.0230 &           0.974 $\pm$          0.338 &          0.3384 $\pm$         0.0521 &          0.3302 $\pm$         0.0571 &           ... &           0.698 $\pm$          0.048 \\
                    Only K2 Primary &         0.00090 $\pm$        0.00086 &          88.850 $\pm$          0.150 &          0.6381 $\pm$         0.0089 &           0.739 $\pm$          0.155 &          0.3690 $\pm$         0.0287 &          0.2726 $\pm$         0.0347 &           0.641 $\pm$          0.315 &           ... \\
                  Only K2 Secondary &         0.00142 $\pm$        0.00116 &          88.809 $\pm$          0.252 &          0.6324 $\pm$         0.0203 &           0.797 $\pm$          0.251 &          0.3489 $\pm$         0.0359 &          0.2811 $\pm$         0.0528 &           0.735 $\pm$          0.135 &           ... \\

\enddata
\end{deluxetable}

   \begin{figure*}
 \epsscale{1.12}
 \includegraphics[scale=0.51,trim=0.0cm 12.0cm 0.0cm 0.0cm]{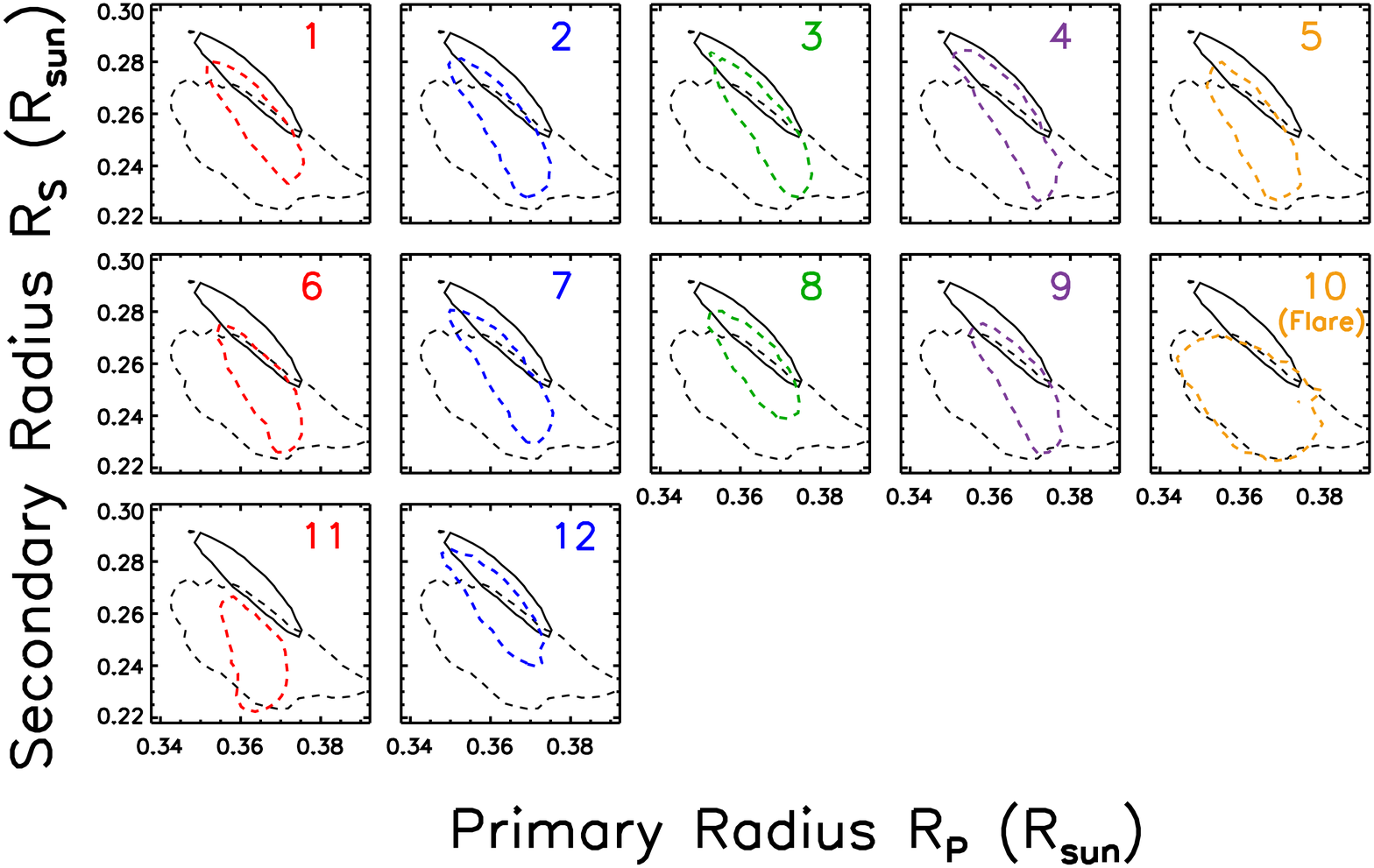}
 \caption{Joint credible intervals (drawn at 68.7\%) on the stellar radii of \ptfeb A+B, as derived from subsets of the data. We show the intervals when fitting the \Ktwo light curve, spectroscopic prior, and RVs (thick solid black contours), only the \Ktwo secondary eclipses, spectroscopic prior, and RVs (dashed black contours), and the credible intervals using only one \Ktwo primary eclipse, the \Ktwo secondary eclipses, spectroscopic prior, and RVs (dashed color contours). Each panels shows the result for using only one primary eclipse, and the single-eclipse fits are color coded to denote eclipses that occulted the same range of longitudes on the primary star: \#1/6/11 (red), \#2/7/12 (blue), \#3/8 (green), \#4/9 (purple), and \#5/10 (orange). Eclipse \#10 occurred just after a flare, so we do not include it in any analysis. \label{fig:radfig}} 
  \end{figure*}
  
     \begin{figure*}
 \epsscale{1.12}
 \includegraphics[scale=0.84,trim=0.0cm 0.0cm 0.0cm 0.0cm]{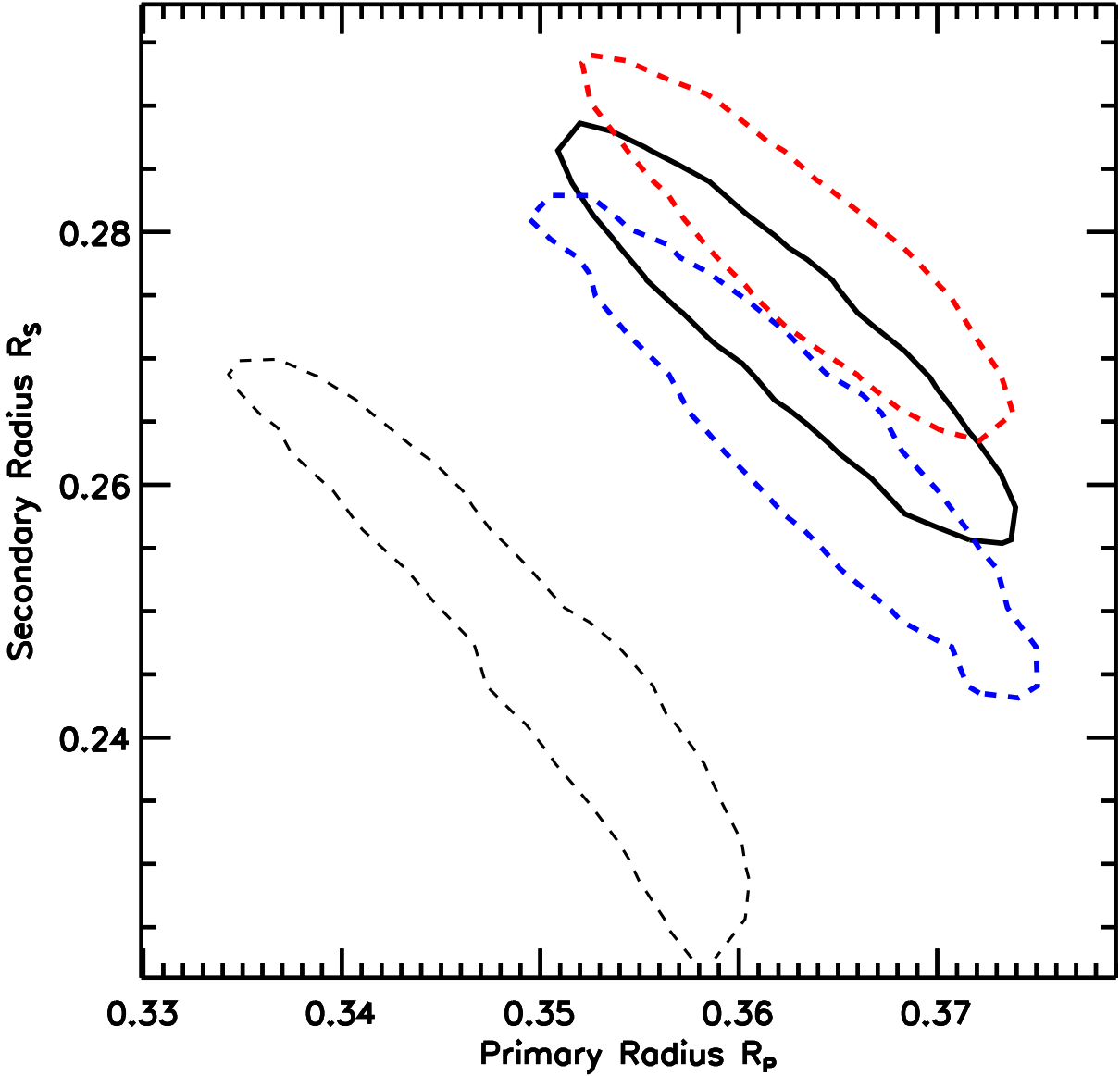}
 \caption{Joint credible intervals (drawn at 68.7\%) on the stellar radii of \ptfeb A+B, as derived using different assumptions about the stellar limb darkening. The solid black contour shows the results for our adopted limb darkening parameters, while the red and blue dashed contours represent a change in the linear coefficient of $\Delta u_1 = 0.1$ upward or downward, as is typically seen between different theoretical treatments of limb darkening (e.g., \citealt{Claret:2011aa} vs \citealt{Sing:2010aa}). The black thin dashed contour shows a fit with no limb darkening, and hence treating the stars as uniformly illuminated disks. The red and blue contours show that for typical variations in adopted limb darkening parameters, the resulting systematic uncertainty in the sum of the radii and the individual radii is $\sim$1--2\%. For this specific system, most of the impact from small changes in limb darkening is reflected in the radius of the secondary. Even in the worst-case scenario of using no limb darkening, the stars are only 6\% smaller, with similar impact on both stars. We therefore conclude that differences in the detailed treatment of limb darkening can not explain the observed scatter in the field mass-radius relation ($\pm 5\%$) or the radius discrepancy we see for the secondary star (20\%). \label{fig:LDfig}} 
  \end{figure*}

   \begin{figure*}
 \epsscale{1.15}
 \plotone{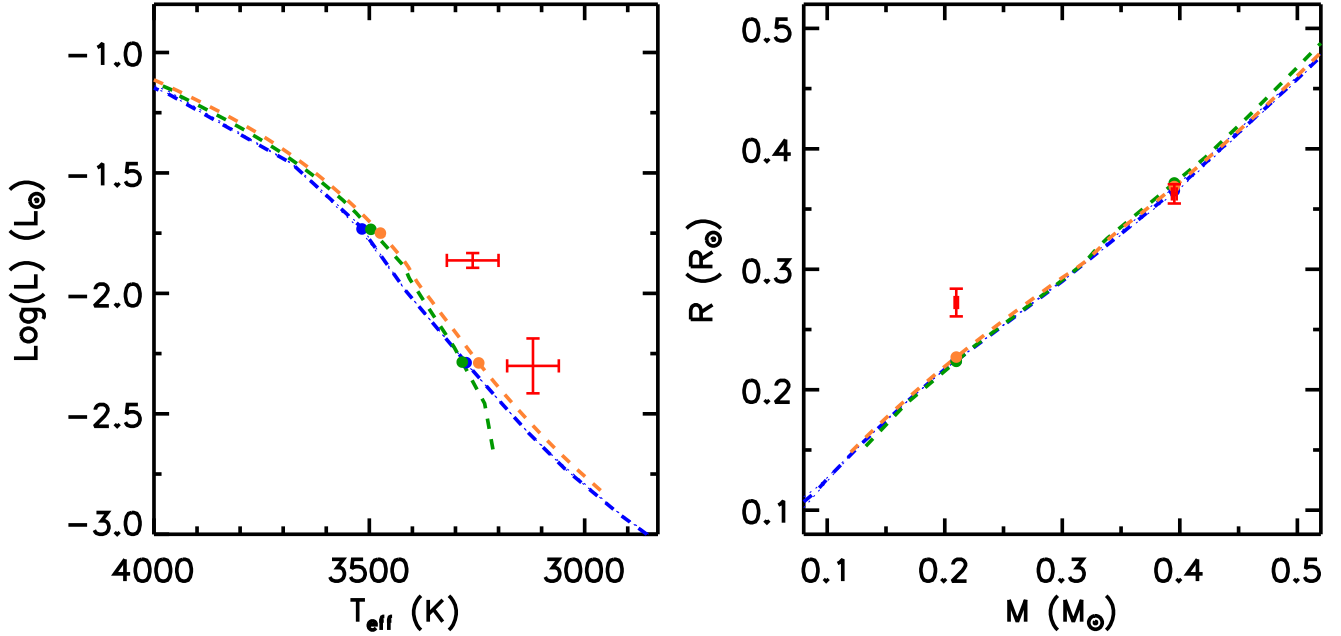}
 \caption{Left: $L$-$T_{\rm eff}$ HR diagram showing our spectroscopic measurements for the components of \ptfeb (red) and the theoretical isochrones at $\tau = 600$ Myr for BHAC15 (blue), DSEP with $[Fe/H]=0$ (green), and DSEP with $[Fe/H]=0.14$ (orange). The interpolated model positions for our measured masses of the two stars are shown with filled circles. The models predict substantially higher temperatures ($\Delta T_{eff} \sim 150$--250 K) than those that result from our spectroscopic analysis or from the geometric $T_{eff}$ corresponding to our radius measurements. Right: Mass-Radius diagram showing the components of \ptfeb and the predictions of the BHAC15 and DSEP models at $\tau = 600$ Myr. For both model sets, the primary agrees well with the mass-radius relations, but the secondary star's radius is significantly underpredicted ($\Delta R \sim 20$\%) by theory. For both the HR diagram and mass-radius diagram, metallicity does not change the model tracks by a sufficient amount to explain the observed discrepancies. The BHAC15 isochrones for $\tau = 500$ Myr and $\tau = 1$ Gyr fall underneath the $\tau = 625$ Myr isochrone in these plots, so uncertainty in the age of Praesepe (e.g., \citealt{Brandt:2015fv}) also can not explain the discrepancies. ( \label{fig:hrdmrd}} 
  \end{figure*}

  \floattable
\begin{deluxetable}{lr}
\tabletypesize{\footnotesize}
\tablewidth{0pt}
\tablecaption{System Parameters for GU Boo Based on Data from LMR05 \label{tab:GUpartable}}
\tablehead{}
\startdata
\hline \multicolumn{2}{l}{\it Orbital Parameters}\\ \hline
                          $T_0$ (HJD) &   2452723.98150 $\pm$        0.00006 \\
                          $T_P$ (HJD) &   2452723.98150 $\pm$        0.00006 \\
                           $P$ (days) &       0.4887279 $\pm$      0.0000008 \\
                             $a$ (AU) &        0.012934 $\pm$       0.000009 \\
                                  $e$ &         0 \\
                            $i$ (deg) &          87.464 $\pm$          0.075 \\
                       $\omega$ (deg) &         270\\
                      $\gamma$ (km/s) &          -24.25 $\pm$           0.11 \\
\hline \multicolumn{2}{l}{Stellar Bulk Parameters}\\ \hline
              $M_p+M_s$ ($M_{\odot}$) &          1.2084 $\pm$         0.0024 \\
                        $q = M_s/M_p$ &           0.981 $\pm$          0.002 \\
                  $M_p$ ($M_{\odot}$) &          0.6099 $\pm$         0.0017 \\
                  $M_s$ ($M_{\odot}$) &          0.5985 $\pm$         0.0010 \\
              $R_p+R_s$ ($R_{\odot}$) &          1.2354 $\pm$         0.0032 \\
                            $R_s/R_p$ &           1.051 $\pm$          0.021 \\
                  $R_p$ ($R_{\odot}$) &          0.6023 $\pm$         0.0068 \\
                  $R_s$ ($R_{\odot}$) &          0.6332 $\pm$         0.0060 \\
\hline \multicolumn{2}{l}{Stellar Atmospheric Parameters}\\ \hline
                    $S_{s,K}/S_{p,K}$ &           0.880 $\pm$          0.006 \\
                    $S_{s,P}/S_{p,P}$ &           0.868 $\pm$          0.007 \\

\enddata
\end{deluxetable}

 \begin{figure*}
 \epsscale{1.12}
  \includegraphics[scale=0.8,trim=0.0cm 0.0cm 0.0cm 0.0cm]{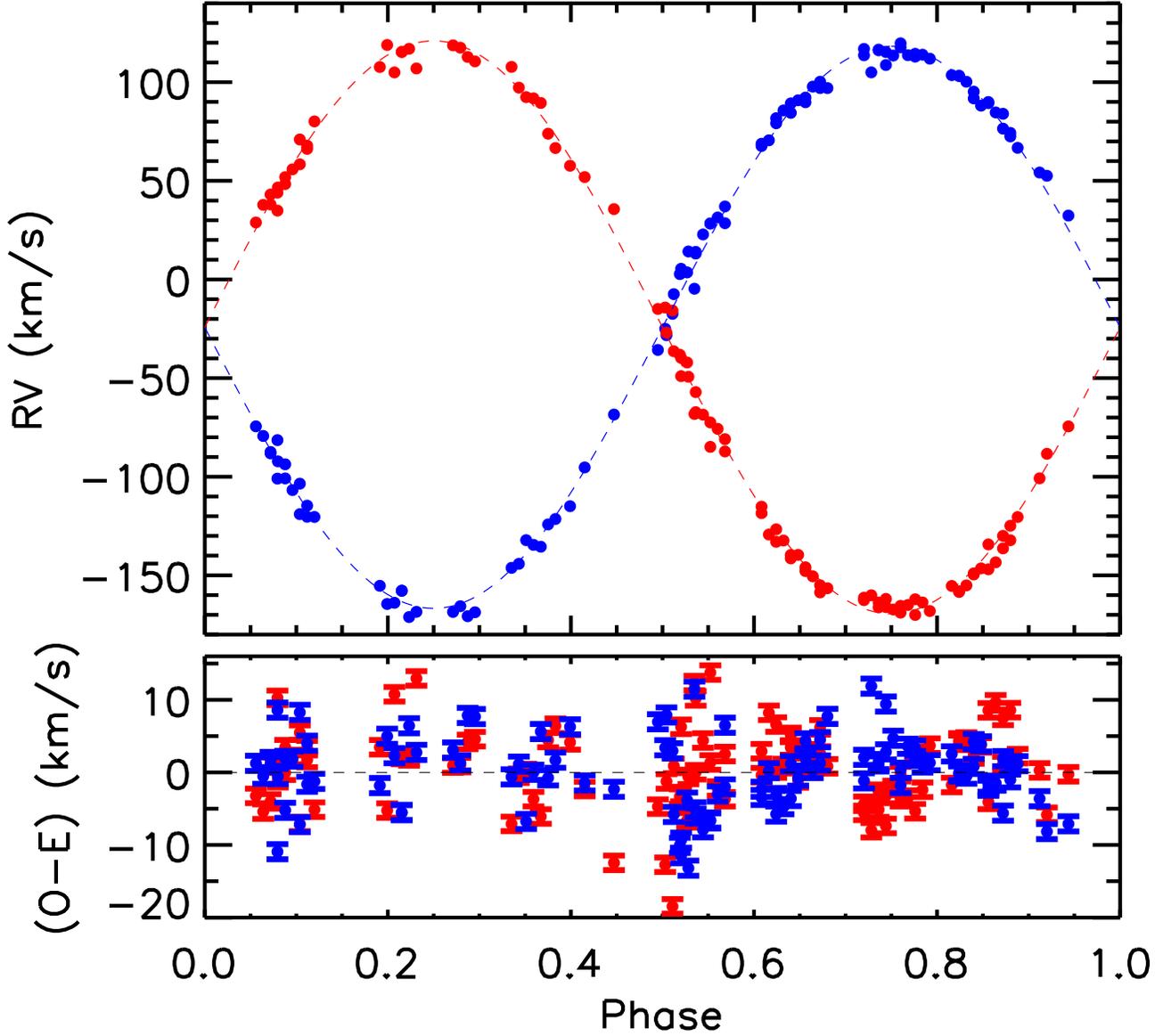}
 \caption{Radial velocities $v_p$ (blue) and $v_s$ (red) for the primary and secondary stars of GU Boo, as reported by LMR05. We also show the best-fit model as determined from our fitting procedure (Appendix A). Underneath, we show the (O-E) residuals with respect to the best-fit model. \label{fig:GUrvfit}} 
  \end{figure*}

   \begin{figure*}
 \epsscale{1.12}
% \plotone{guboo/Rplot1110.eps}
 \plotone{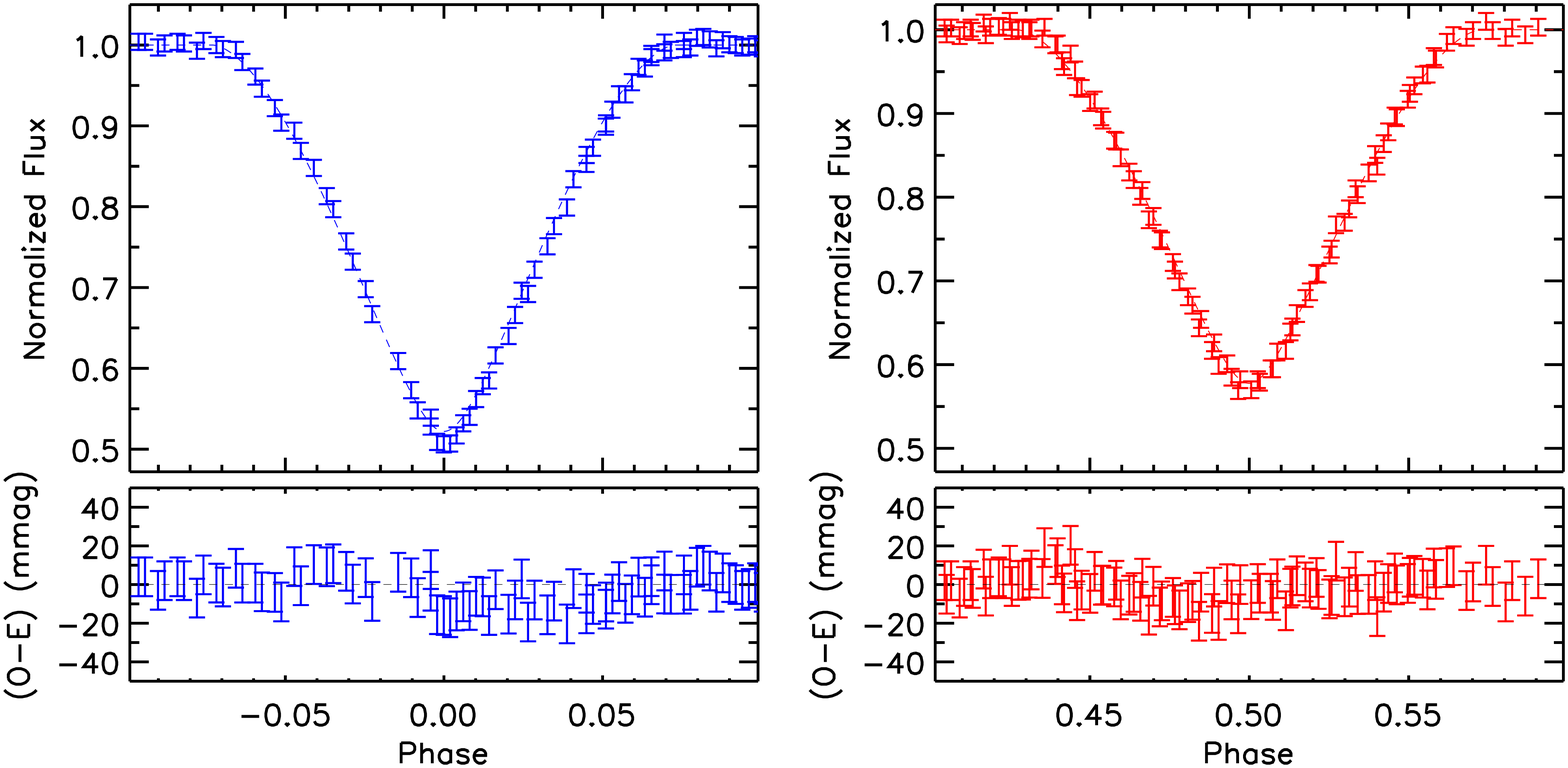}
 \caption{$R$ band photometry for the primary eclipse (left) and secondary eclipse (right) of GU Boo, as reported by LMR05, along with the best-fitting models (dashed lines) and the (O-E) residuals (bottom panels). \label{fig:GURfit}} 
  \end{figure*}

     \begin{figure*}
 \epsscale{1.12}
% \plotone{guboo/Iplot1110.eps} 
 \plotone{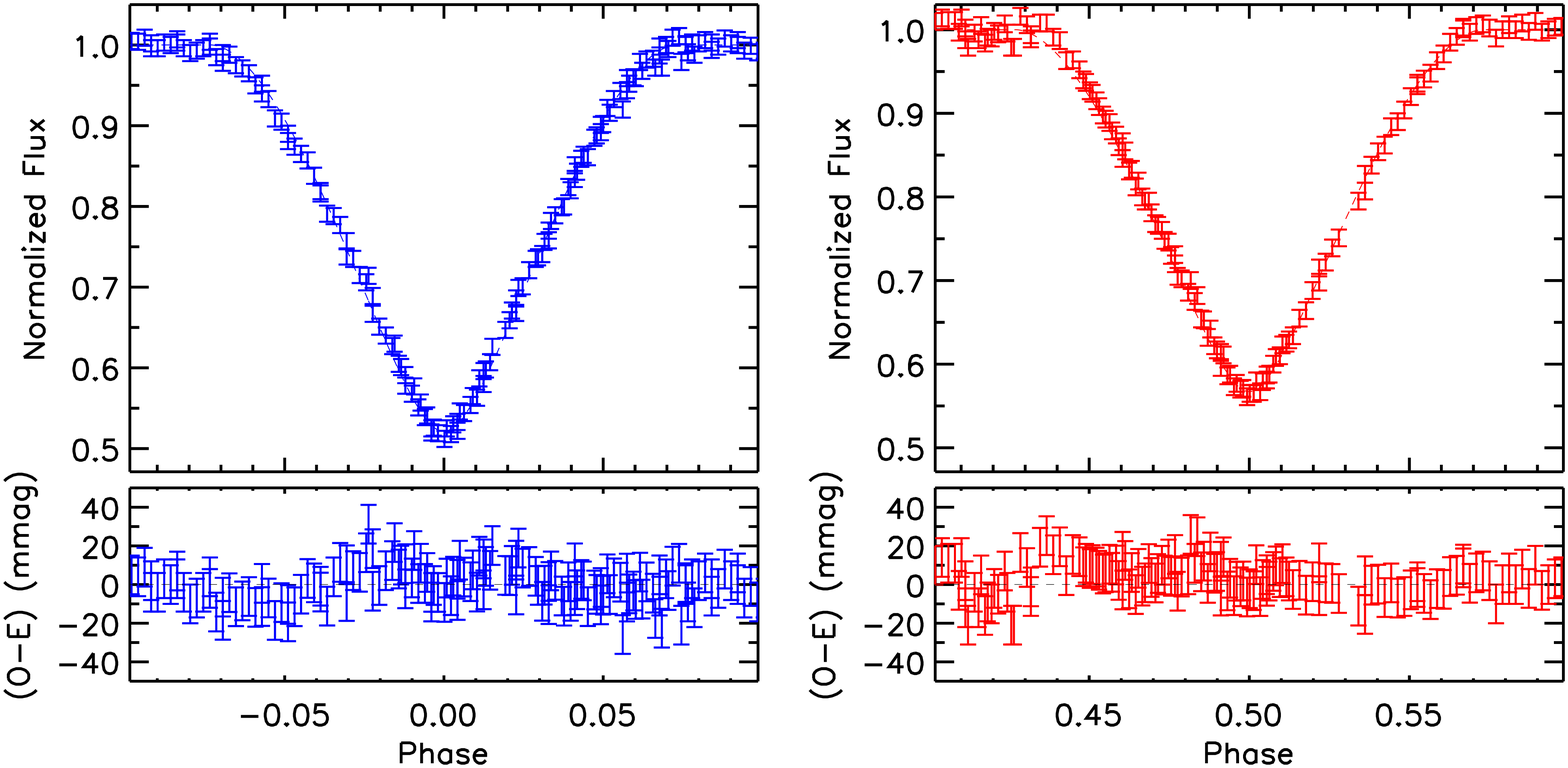}
 \caption{$I$ band photometry for the primary eclipse (left) and secondary eclipse (right) of GU Boo, as reported by LMR05, along with the best-fitting models (dashed lines) and the (O-E) residuals (bottom panels). \label{fig:GUIfit}} 
  \end{figure*}

   \begin{figure*}
 \epsscale{1.12}
 \includegraphics[scale=0.44,trim=0.0cm 0.0cm 0.0cm 0.0cm]{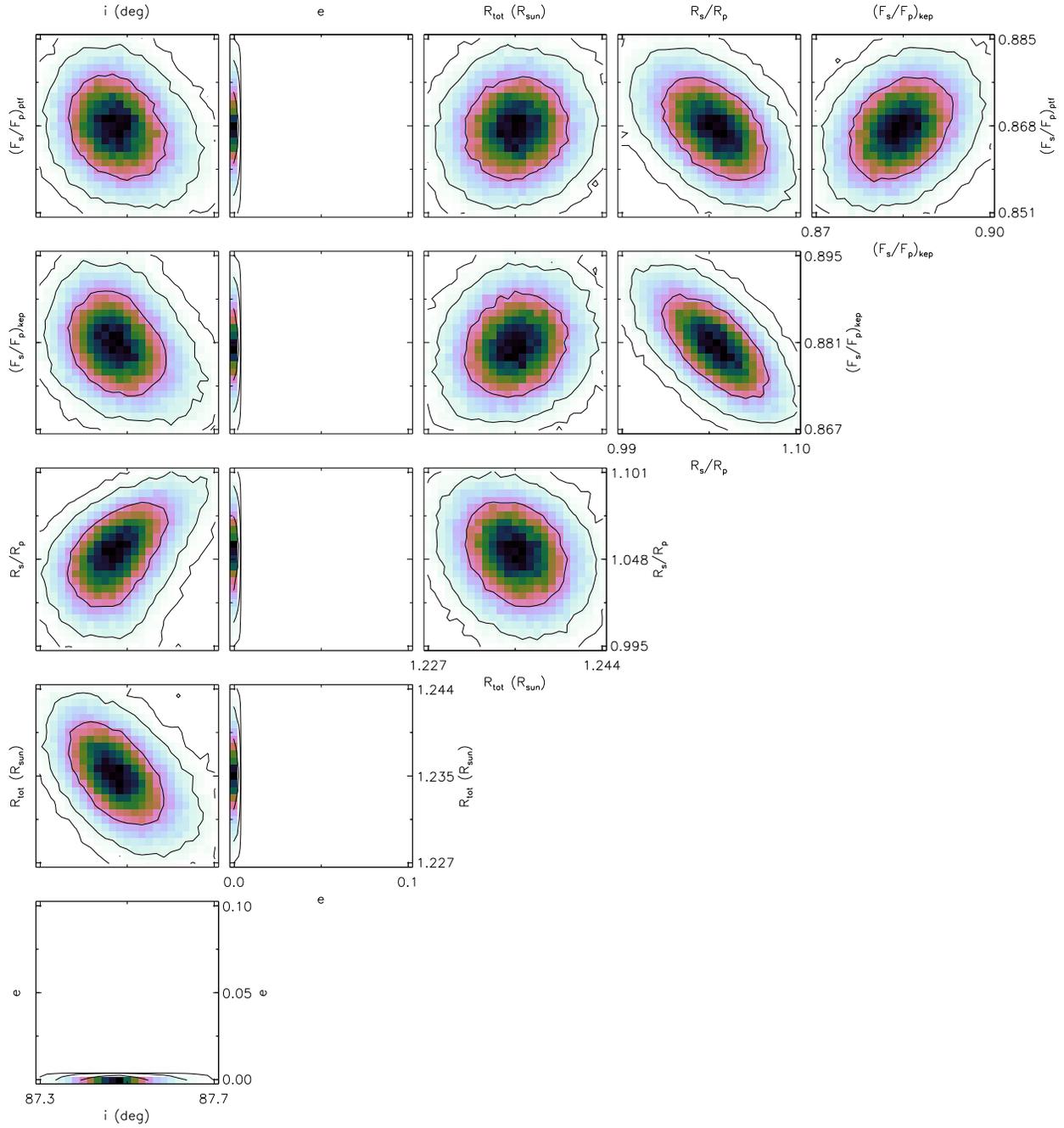}
 \caption{Two-dimensional marginalized posterior distributions for the 6 fit parameters in our MCMC analysis of GU Boo that are most likely to show astrophysically important covariances. There are moderate covariances between the radius ratio, radius sum, and system inclination, (column 1, rows 3--4) as well as between the radius ratio and the surface brightness ratios (column 4, row 2 and column 5, row 1). \label{fig:GUcorner}} 
  \end{figure*}

\end{document}